\documentclass[%
 reprint,
 amsmath,amssymb,
 aps,
]{revtex4-2}

\usepackage{graphicx}
\usepackage{dcolumn}
\usepackage{bm}
\usepackage{xcolor}
\usepackage{float}

\begin{document}

\preprint{APS/123-QED}
\title{A self-consistent refractive index model for fast simulation of free-electron lasers}

\author{River R. Robles}
\email{riverr@stanford.edu}
\author{Gabriel Marcus}
\author{Zhirong Huang}%
\affiliation{%
 SLAC National Accelerator Laboratory, Menlo Park, CA 94025.
}%

\date{\today}

\begin{abstract}
Modern x-ray free-electron lasers (XFELs) produce x-ray pulses of exceptional transverse coherence. This is due largely to the process of optical guiding by which the radiation is both refractively guided by the bunched electron beam and gain guided by the preferential amplification of on-axis radiation. These effects may be summarized by an effective index of refraction, which has been used in the past to study the transverse dynamics of the FEL process with significant simplifications and approximations, but never fully self-consistently. We present here a self-consistent method for studying high gain FELs in the linear regime by approximating the FEL equations to second-order in the lateral displacement from the nominal electron beam axis. This is made possible by casting the FEL equations in the language of optical fibers with an appropriately chosen refractive index. We demonstrate that this approach is both fast and highly accurate, indicating that the most important FEL dynamics are inherently second-order. In its full form our method can capture the effects of transverse offsets in both the x-ray beam and the electron beam, making it a versatile tool for studying non-ideal effects in seeded FELs, regenerative amplifier XFELs, and even self-amplified spontaneous emission (SASE) FELs. 

\end{abstract}

\maketitle

\section{\label{sec:intro}Introduction}

X-ray free-electron lasers provide the x-ray user community access to x-ray pulses of incredible brightness and transverse coherence. The predominant reason for the excellent quality of the output of the XFEL is the phenomenon of optical guiding \cite{scharlemann1985optical}, by which the radiation beam is both refractively guided and amplified by the electron beam. Optical guiding guarantees that the output radiation is dominated by a well-focused, nearly gaussian transverse mode which is by definition well-matched to the electron beam and is therefore strongly amplified along the undulator length \cite{moore1985high,huang2007review}.

The equations which govern the XFEL have been well-studied and validated against experimental results. Unfortunately, except in very restricted approximate situations, these equations can only be studied by complicated numerical codes (most notably, Genesis \cite{reiche1999genesis}). This has two significant drawbacks. First, these codes can simply be time-consuming since they involve self-consistently tracking both the phase space variables of the macroparticles representing the electron beam, as well as the radiation profile. This is especially true in time-dependent simulations of self-amplified spontaneous emission (SASE), which is inherently a statistical process which requires many independent simulations to properly characterize the properties of the output radiation. The lengthy nature of FEL simulations can even be prohibitive when time-independent simulations suffice, such as in the tolerance studies associated with a regenerative amplifier x-ray FEL (RAFEL) \cite{huang2006fully,marcus2020refractive} where the field must be tracked on a large, well-resolved transverse grid over many passes through an optical cavity.

The second drawback to the standard approach to XFEL studies is that numerical simulations of these very complicated processes can often be difficult to interpret physically, and the equations themselves are complicated enough that it is difficult to gain any analytic understanding if even one non-ideal feature is present. For example, all XFELs are affected by some level of jitter in the trajectory of the driving electron beam, which can lead to degradation of the eventual x-ray pulse energy and pointing jitter of the x-ray beam since most of the final energy in the pulse is developed in the last gain length of the undulator. Some simpler and faster numerical methods have been proposed which can handle these non-ideal effects, however they still implement relatively opaque numerical methods, and interpreting their results correctly demands some nuance which we will discuss here \cite{baxevanis2017}.

In this paper, we present an attempt at rectifying these issues by introducing a fast numerical scheme for studying XFELs which is both accurate and rooted in a simple physical understanding of the FEL process based on optical guiding. Our approach is based on interpreting the three-dimensional FEL equations as those of an optical fiber with appropriately chosen variable parabolic refractive index \cite{marcuse2013}, and further approximating the transverse radiation mode as a simple gaussian of variable size and centroid. We note that interpreting the FEL interaction as being described by an effective index of refraction is not a new idea \cite{scharlemann1985optical, sprangle1987analysis, sprangle1987radiation, prosnitz1981high}, but this is, to the authors' knowledge, the first self-consistent implementation of such a method, and additionally the first implementation rooted in a fully three-dimensional formulation of the FEL equations including all non-ideal effects: most importantly emittance, energy spread, diffraction, and orbit errors. This approach has the significant advantage that the expensive process of tracking both the phase space variables of the macroparticles in the electron beam and the radiation field on a large transverse grid is replaced by tracking just a handful of complex numbers which characterize the gaussian radiation field and the effective refractive index of the electron beam. Finally, the accuracy of this method highlights that the most important physics involved in the optical guiding process can be captured with a simple second-order model, thereby paving the way towards even simpler approaches. 

The structure of the rest of the paper is as follows. In Section \ref{sec:approach} we introduce the numerical scheme, reviewing both how radiation fields propagate in quadratic optical fibers and also discussing how the electron beam in an FEL may be treated as such a fiber. In Section \ref{sec:benchmarks} we present comparisons of the present method to the results produced by the standard FEL code Genesis for a variety of representative physical scenarios. Finally in Section \ref{sec:jitter} we employ our method to understand the ramifications of electron beam trajectory errors.

\section{\label{sec:approach}Theoretical Framework}

It is well-established that the fundamental transverse mode of the FEL driven by a gaussian electron beam is itself roughly gaussian (see e.g. \cite{kim2017} and the references therein). This mode has the highest gain in the exponential gain regime and therefore dominates over higher-order Laguerre-Gaussian-like modes. Although in self-amplified spontaneous emission FELs many modes compete with each other in the lethargy regime, for a FEL amplifier seeded above the shot noise power the field is roughly gaussian throughout the undulator up to saturation, and indeed in all standard, high-gain FEL scenarios the gaussian mode is dominant by the undulator exit. Thus we will simplify the problem of tracking the FEL dynamics by assuming the field to take a gaussian form
\begin{equation}\label{eqn:gaussianansatz}
    E(x,y,z) = f(z)\exp\left[-\frac{i}{2}\left(Q_x(z)(x-x_0(z))^2+Q_y(z)y^2\right)\right].
\end{equation}
By using this ansatz, we simplify the problem by reducing the number of required parameters to four. These parameters $f(z)$, $Q_x(z)$, $Q_y(z)$, and $x_0(z)$, describe the field amplitude, mode size and divergence in x and y, and transverse centroid, respectively. In general they are all complex. In Subsection \ref{subsec:beamsinfibers} we will discuss how a beam of this form behaves as it propagates through a general optical fiber with a second-order transverse index gradient. In Subsection \ref{subsec:felindex} we will start from the FEL wave equation to derive the appropriate second-order form of the FEL refractive index, and the tracking of the electron beam phase space variables is simplified to tracking the four components of the refractive index at each integration step. 

\subsection{\label{subsec:beamsinfibers}Gaussian beams in quadratic optical fibers}

The radiation profile in a dielectric waveguide with some spatially varying refractive index $n(r,z)$ satisfies the following paraxial wave equation (see e.g. \cite{marcuse2013})
\begin{equation}\label{eqn:dielectricwaveguideeqn}
    2ik_r\frac{\partial E}{\partial z} + \nabla_\perp^2E = k_r^2\left(1-n(r,z)^2\right)E.
\end{equation}
We will show in the next section how the effective refractive index of the FEL comes about and how it may be written in the following general form to lowest-order:
\begin{equation}\label{eqn:secondorderindex}
    n(x,y,z)^2 = n_0(z)^2+2n_1(z)x-n_{2x}(z)x^2-n_{2y}(z)y^2.
\end{equation}
The presence of the linear-order term allows us to accurately take into account radiation profiles with non-zero transverse centroid, where the direction x is defined by the direction in which the centroid is offset, without loss of generality. This index couples naturally to our gaussian ansatz for the radiation field, as we can see from plugging the gaussian ansatz into Equation \ref{eqn:dielectricwaveguideeqn}. Since operating on a gaussian with the transverse laplacian operator returns the same gaussian multiplied by a polynomial in x and y, we may extract differential equations for the four mode parameters by matching coefficients of like powers of the transverse coordinates in this equation. The result is the following four first-order differential equations:
\begin{equation}
\label{eq:tracking}
\begin{split}
    Q_x'(z) =& \frac{k_r^2n_{2x}(z)+Q_x(z)^2}{k_r},\\
    Q_y'(z) =& \frac{k_r^2n_{2y}(z)+Q_y(z)^2}{k_r},\\
    x_0'(z) =& \frac{k_r(n_1(z)-n_{2x}(z)x_0(z))}{Q_x(z)},\\
    f'(z) =& \frac{f(z)}{2k_r}\left[Q_x(z)+Q_y(z)+ik_r^2(n_0(z)^2-1\right.\\
    &\left.+2n_1(z)x_0(z)-n_{2x}(z)x_0(z)^2)\right].
\end{split}
\end{equation}
We note that this approach is analogous to that taken in \cite{kogelnik1965propagation} to study gaussian beam propagation through azimuthally symmetric gradient index fibers. In general, these equations may be integrated through the undulator, given that we understand how to compute the refractive index components $n_0^2$, $n_1$, $n_{2x}$, and $n_{2y}$. Equation \ref{eq:tracking} elucidates the primary reason for the fast nature of the method: we have reduced the problem of tracking the radiation profile down to tracking four complex numbers. 

Before moving on, we note that these equations present a simple method for analyzing the high-gain regime of the FEL. The steady-state high-gain regime is characterized by a constant beam size, on-axis propagation, and exponential gain in the field strength. We thus expect $Q(z)=Q_{hg}$, $f(z)=f_0e^{i\mu_{hg} z}$, and $x_0(z)=0$, where $Q_{hg}$ is the high-gain mode parameter and $\mu_{hg}$ is the high-gain growth rate. In this regime, the refractive index components are similarly constant, and satisfy equations which may in general be solved for the high-gain growth rate and mode size parameter:
\begin{equation}
    \begin{split}
        n_{2,hg} &= -\frac{Q_{hg}^2}{k_r^2},\\
        n_{0,hg}^2 &= 1 + \frac{2}{k_r}\left(\mu_{hg}+\frac{iQ_{hg}}{k_r}\right).
    \end{split}
\end{equation}
These equations, once the refractive index component forms are specified, constitute a dispersion relationship analogous to that from standard FEL theory which can be solved to obtain the high-gain mode size parameter and growth rate. Knowledge of the form of the refractive index components will come directly from the FEL equations in the next section. 

\subsection{\label{subsec:felindex}The FEL as a quadratic optical fiber}

We will borrow from the fully three-dimensional FEL formalism developed in \cite{baxevanis2017} in order to understand how we may model the FEL as an optical fiber. As such, we consider here a planar undulator without any taper. Those authors showed that the FEL field development could be written concisely in the form of a single integro-differential equation
\begin{equation}
\label{eqn:panoseqn}
\begin{split}
    \frac{\partial E}{\partial z}&+\frac{1}{2ik_r}\nabla_\perp^2E = \\
    &\int dp_xdp_y\int_0^zd\zeta K_1(x,y,p_x,p_y,z,\zeta)E(x_+,y_+,\zeta),
\end{split}
\end{equation}
where $x_+=x\cos(k_\beta(\zeta-z))+(p_x/k_\beta)\sin(k_\beta(\zeta-z))$ with a similar expression in y. Furthermore, $k_\beta$ is the wavenumber associated with the beam betatron oscillations which we assume for now to be matched to an external smooth focusing lattice such that $\sigma_{x'}=k_\beta\sigma_x$, and the beam size and angle are symmetric between the two planes and constant along the undulator. Additionally, the integral kernel $K_1$ has the form
\begin{widetext}
\begin{equation}
    K_1(x,y,p_x,p_y,z,\zeta)=K_{10}(\zeta-z)\exp\left[-\frac{(p_x^2+p_y^2+k_\beta^2x^2+k_\beta^2y^2)}{2k_\beta^2}\left(\frac{1}{\sigma_x^2}+ik_rk_\beta^2(\zeta-z)\right)\right].
\end{equation}
\end{widetext}
In this expression,
\begin{equation}
    K_{10}(\xi) = -\frac{8i\rho^3k_u^3}{2\pi k_\beta^2\sigma_x^2}\xi\exp[-i\Delta\nu k_u\xi-2\sigma_\delta^2k_u^2\xi^2],
\end{equation}
where $\rho$ is the Pierce parameter \cite{bonifacio1984collective}, $k_u$ is the undulator wavenumber, $\sigma_\delta$ is the relative energy spread, and $\Delta\nu$ is the FEL detuning parameter. Identifying the right-hand side of Equation \ref{eqn:panoseqn} with that of Equation \ref{eqn:dielectricwaveguideeqn} we may extract the effective local refractive index of the FEL 
\begin{equation}
\begin{split}
    n(x,y,z)^2 =& 1-\frac{2i}{k_r}\frac{1}{E(x,y,z)}\int dp_xdp_y\\
    &\times\int_0^zd\zeta K_1(x,y,p_x,p_y,z,\zeta)E(x_+,y_+,\zeta).
\end{split}
\end{equation}
In Appendix \ref{app:index} we present an analytic form for this index for the gaussian field ansatz after the angular integrals have been taken, including the effects of the modifications we will present in Section \ref{sec:jitter}. For the purposes of this work we would like to choose an approximation to put this index in the form of Equation \ref{eqn:secondorderindex}. Although there are in principle many ways to do this, for now we will consider a simple on-axis Taylor expansion. Physically, we expect that this model will capture most of the relevant physics since the gain medium, the electron beam, is on axis. Later on, when we consider errors in the electron orbit, we will opt instead to expand around the local electron beam centroid for the same reason. For radiation which is seeded far off-axis one expects that this will yield poor agreement in the early sections of the undulator, but, as long as there is sufficient total gain along the undulator length, we should expect good agreement by the undulator exit. The accuracy of this relatively simple approach will be established more quantitatively in the next section through numerical benchmarks. Thus we assign the approximate coefficients by 
\begin{equation}
\begin{split}
    n_0(z)^2 &= n(x,y,z)^2\bigg\rvert_{x=y=0},\\
    n_1(z) &= \frac{1}{2}\frac{\partial n^2}{\partial x}\bigg\rvert_{x=y=0},\\
    n_{2x}(z) &= -\frac{1}{2}\frac{\partial^2n^2}{\partial x^2}\bigg\rvert_{x=y=0},\\
    n_{2y}(z) &= -\frac{1}{2}\frac{\partial^2n^2}{\partial y^2}\bigg\rvert_{x=y=0}.
\end{split}
\end{equation}
These expressions can be evaluated explicitly as a function of the local mode parameters such that the only integral left over is the integral in $\zeta$ which can be completed numerically in general. These, coupled with the results of Subsection \ref{subsec:beamsinfibers}, provide a self-consistent numerical scheme for propagating the seed radiation through the FEL which demands the tracking of just eight complex variables.

\subsection{\label{subsec:physicalvalues}Conversion to physical values}

To facilitate comparison of our results with simulations we must convert the four complex radiation mode parameters we are tracking into measurable quantities. The parameters of the most interest are the radiation rms or fwhm size, physical centroid, angular centroid, and power. The fwhm we will discuss is that of the norm of the transverse field profile $\sqrt{EE^*}$ \footnote{This choice is arbitrary in the sense that all comparisons could equivalently be made to the intensity profile $EE^*$.}. It is straightforward to show this to be 
\begin{equation}
    \text{fwhm}_{x,y} = 2\sqrt{-\frac{2\log(2)}{\Im[Q_{x,y}]}},
\end{equation}
where $\Im[.]$ extracts the imaginary part of its argument. Since the beam is by definition gaussian the rms size is related to this by fwhm $=2\sqrt{2\log(2)}$rms. Furthermore the physical centroid, which we identify as the location of the peak of the intensity profile, is 
\begin{equation}
    x_{cen} = \Re[x_0]+\Im[x_0]\frac{\Re[Q_x]}{\Im[Q_x]},
\end{equation}
where $\Re[.]$ extracts the real part of its argument. The angular centroid on the other hand is the corresponding centroid of the transverse Fourier transform of the field profile
\begin{equation}
    \mathcal{E}(\phi_x,\phi_y,z) \propto \int E(x,y,z)e^{-ik_r(\phi_xx+\phi_yy)}dxdy.
\end{equation}
It is readily found that the centroid of $\sqrt{\mathcal{E}\mathcal{E}^*}$ is
\begin{equation}
    \phi_{x,cen} = -\frac{\left|Q_x\right|^2\Im[x_0]}{k_r\Im[Q_x]}.
\end{equation}
It is also worth noting that we can invert these centroids to write the original complex centroid parameter in terms of the physical and angular centroids
\begin{equation}
    x_0 = x_{cen}+\frac{k_r\phi_{x,cen}}{Q_x}.
\end{equation}
Finally, the time-averaged radiation power is the usual integral over the transverse intensity profile
\begin{equation}
\begin{split}
    P(z) =& \frac{\epsilon_0c}{2}\int \left|E(x,y,z)\right|^2dxdy\\
    =& \frac{\pi\epsilon_0c\left|f(z)\right|^2}{2\sqrt{(-\Im[Q_x])(-\Im[Q_y])}}\exp\left[-\frac{\left|Q_x\right|^2\Im[x_0]^2}{\Im[Q_x]}\right].
\end{split}
\end{equation}
We note here that there are some ambiguities in the choices of observables that are only relevant when the field deviates from a pure gaussian profile, which is not captured in the analysis presented here. In this scenario, for example, the fwhm and rms transverse sizes are no longer related by a factor of $2\sqrt{2\log(2)}$. Additionally, the peak of the intensity profile no longer coincides with the average x position of the intensity profile, defined by 
\begin{align}
    \langle x\rangle(z) = \frac{\int x\left|E(x,y,z)\right|^2dxdy}{\int \left|E(x,y,z)\right|^2dxdy}.
\end{align}
In general the field profile will not remain perfectly gaussian in simulation for reasons we will discuss later, and thus one must think more carefully about which observable quantities are more important. This is in general specific to the particular application in question. We will discuss the implications of these ambiguities in the next section alongside the numerical benchmarks.

\section{\label{sec:benchmarks}Numerical Benchmarks}

To evaluate the validity of the approximations we have made thus far we will compare the predictions of this method with time-independent (single frequency) Genesis simulations of the LCLS-II-HE style FEL. The relevant parameters for this case are described in Table \ref{tab:lclsiihe}. In order to properly compare the simulation results against the theory, we employ a 500 kW seed power in all simulations to avoid saturation by the end of the simulated undulator length. Additionally, the radiation is initialized at a waist at the undulator entrance. We will study the impact of the transverse size, transverse offset, and initial frequency detuning of the initial seed radiation on the validity of the model. 

\begin{table}[h!]
    \centering
    \begin{tabular}{|c|c|c|c|}
    \hline
        Parameter & Variable & Unit & Value  \\
        \hline
        E-Beam Energy & $\gamma mc^2$ & GeV & 8.003\\
        E-Beam Emittance & $\epsilon_{nx,ny}$ & nm-rad & 350\\
        Energy Spread & $\sigma_\gamma/\gamma$ & $10^{-4}$ & 0.875\\
        Pierce Parameter & $\rho$ & $10^{-4}$& 5.45\\
        E-Beam rms Size & $\sigma_{x,y}$ & $\mu$m & 17\\
        Current & $I$ & kA & 1.5\\
        Undulator Period & $\lambda_u$ & cm & 2.6\\
        Seed Wavelength & $\lambda_{rad}$ & \AA & 1.261\\
        Detuning & $\Delta\nu$ & $10^{-4}$ & -8.13\\
        \hline
    \end{tabular}
    \caption{The relevant parameters for the numerical benchmarks are shown. }
    \label{tab:lclsiihe}
\end{table}

\subsection{On-axis seed}

\begin{figure}[htb]
    \centering
    \includegraphics[width=0.85\columnwidth]{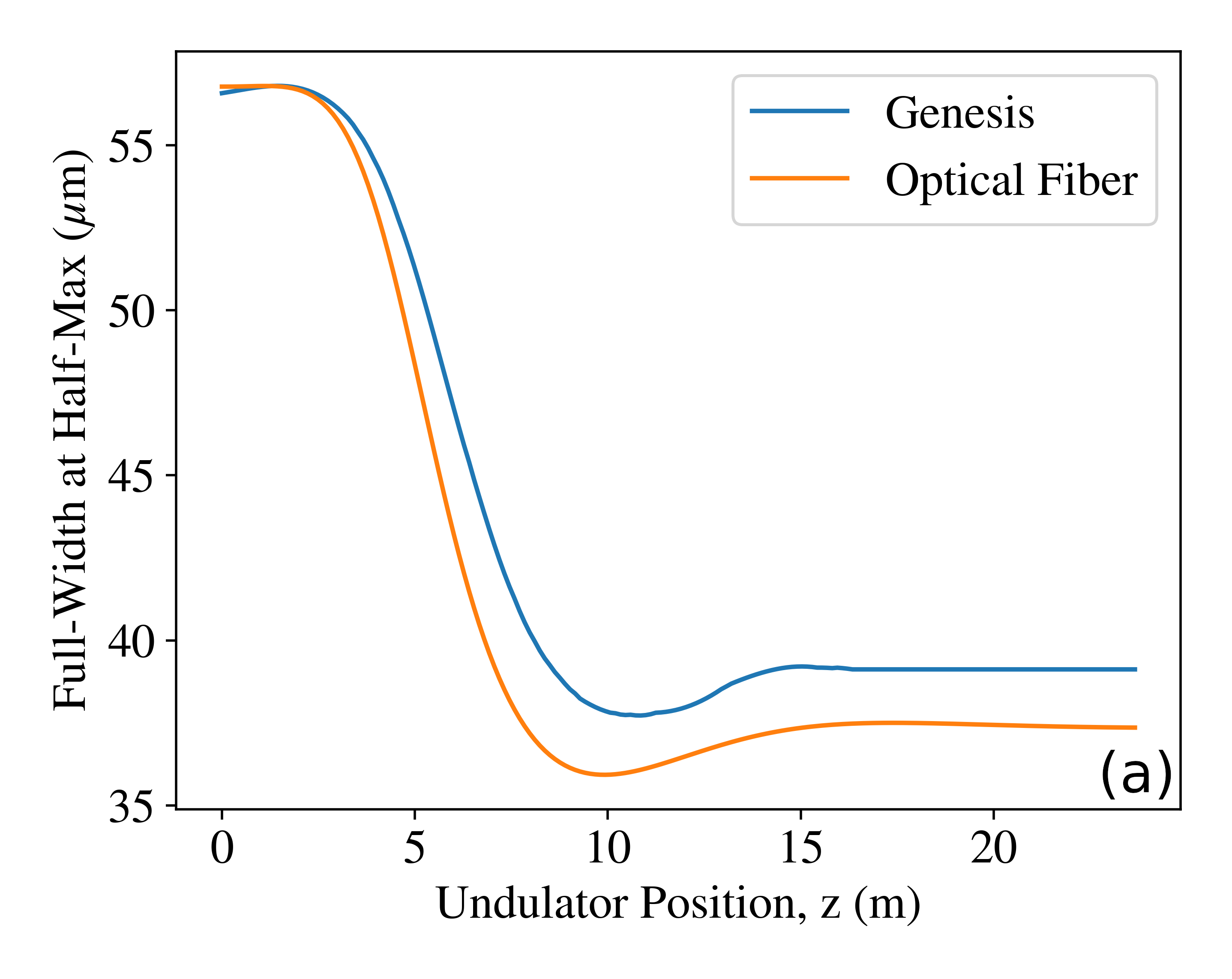}
    \includegraphics[width=0.85\columnwidth]{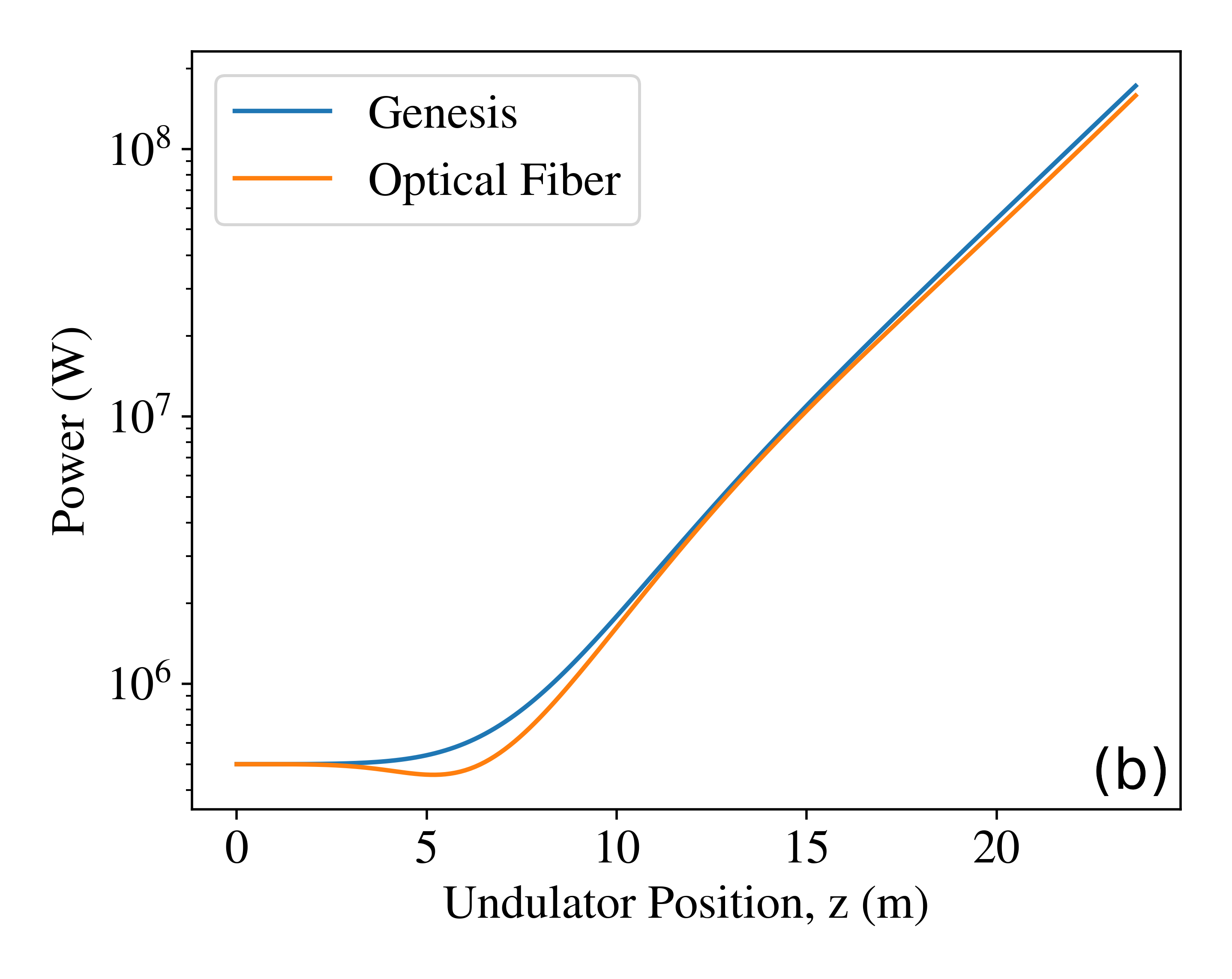}
    \caption{The radiation fwhm (a) and the radiation power growth (b) are plotted against the distance along the undulator from Genesis and from the optical fiber theory for an initial on-axis seed of 29.8 m Rayeligh range.}
    \label{fig:onaxis_benchmarks}
\end{figure}

Before moving to the more general case with an offset in the input seed trajectory, we will establish the validity of our approximate method in an ideal scenario with on-axis seed radiation. We show some representative plots of this scenario in Figure \ref{fig:onaxis_benchmarks}. In this figure we show, on the top, the full-width at half-maximum of the radiation intensity profile, and, on the bottom, the power development through the undulator. This particular case  utilizes a seed radiation field with fwhm of 55 $\mu$m. We observe very good agreement in both metrics. The slight disagreement of the power profile early in the undulator is believed to result from the relative prominence of radiation intensity far off-axis relative  to the electron beam size, as we discuss further in Section \ref{subsec:discussion}.

\begin{figure}[htb]
    \centering
    \includegraphics[width=0.85\columnwidth]{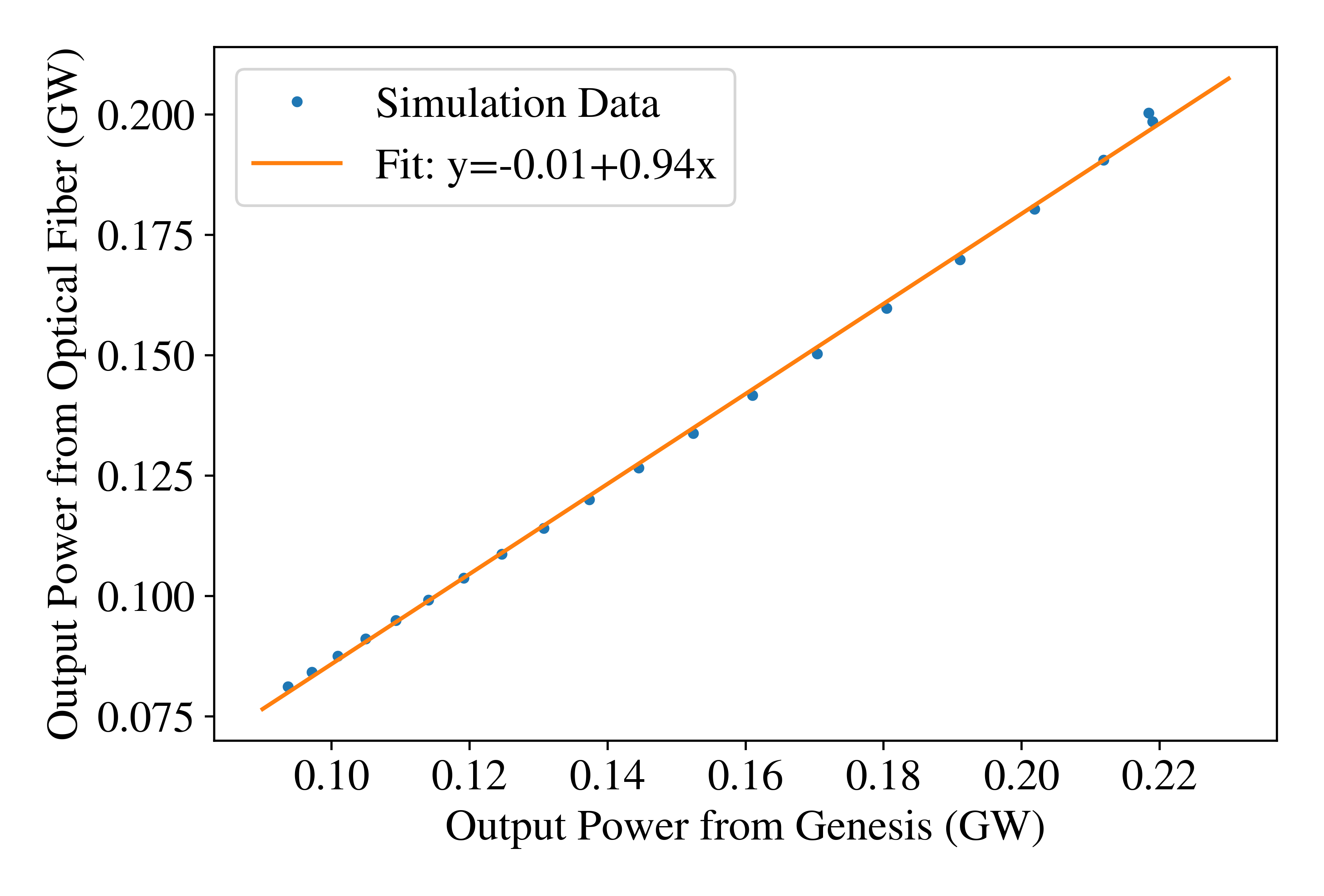}
    \caption{The output power produced from the optical fiber method is plotted against that produced by Genesis as the seed Rayleigh range is scanned from 10 m to 80 m. }
    \label{fig:onaxis_zraylscan}
\end{figure}

To illustrate the accuracy of the method more broadly, we show in Figure \ref{fig:onaxis_zraylscan} a comparison of the output power produced from Genesis and from our optical fiber method while scanning the seed radiation Rayleigh range from 10 m out to 80 m. We observe, as one would hope, a clearly linear trend between the two values with a slope very nearly equal to unity. This implies that the final error in power is relatively systematic across different input seed parameters and therefore does not introduce any extraneous correlations between the final power and the input spot size. As a result, one can trust the approximate results produced by the method across a broad range of input radiation mode sizes despite the fact that the method is explicitly valid only near the axis. This is again a simple result of the relative dominance of the radiation power close to the axis over that far from the axis by the undulator end. 

\begin{figure}[h!]
    \centering
    \includegraphics[width=0.85\columnwidth]{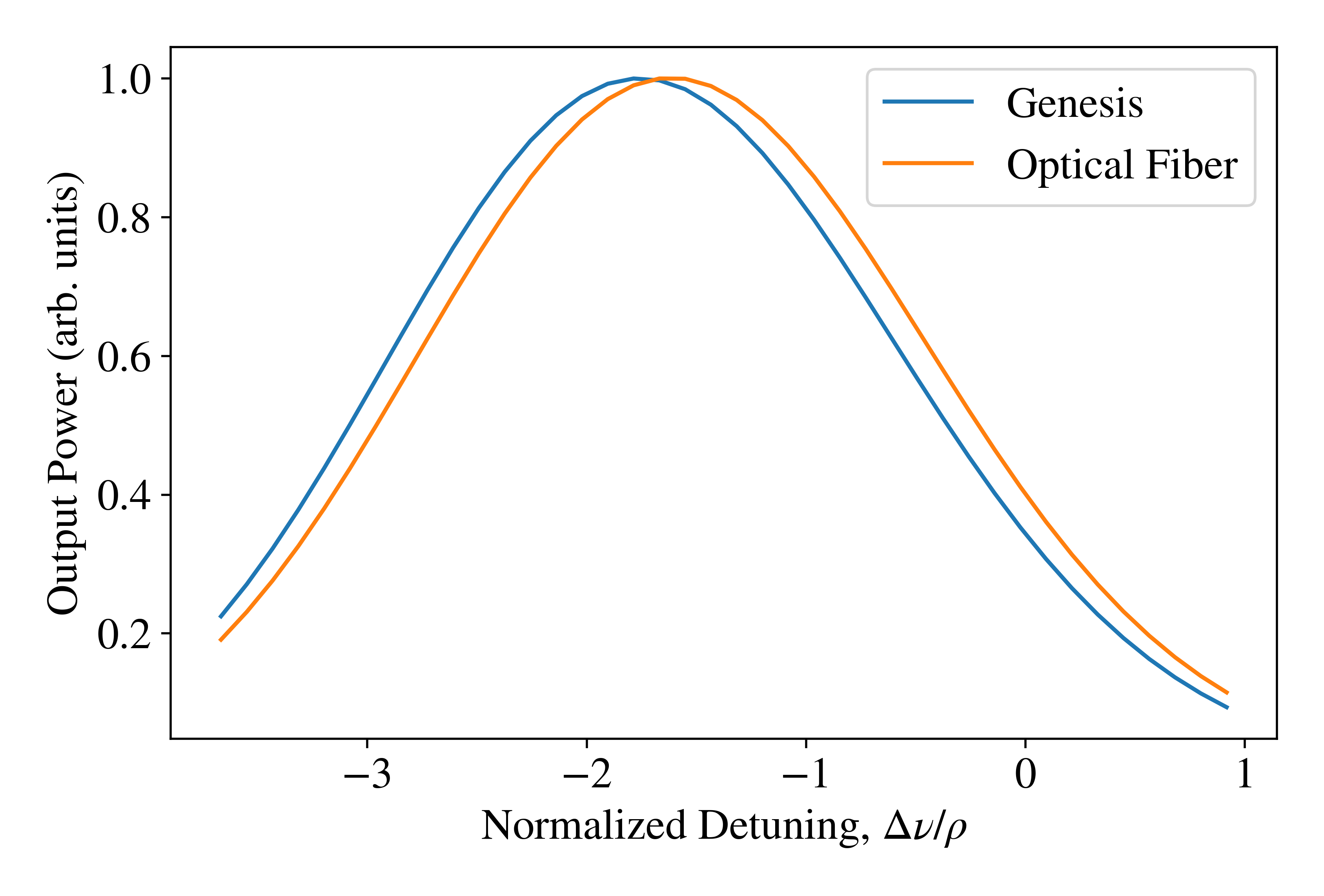}
    \caption{The radiation power gain normalized to the power gain at optimal detuning is plotted as a function of frequency detuning as obtained in Genesis and from the optical fiber method. }
    \label{fig:detuning}
\end{figure}

We also take this opportunity to discuss the validity of this approach across a range of frequency detunings $\Delta\nu=(\lambda_{res}-\lambda_{rad})/\lambda_{rad}$ where $\lambda_{res}$ is the resonant wavelength for the specified electron and undulator parameters, defined as 
\begin{equation}
    \lambda_{res} = \frac{\lambda_u}{2\gamma^2}\left(1+\frac{K^2}{2}\right),
\end{equation}
and $\lambda_{rad}$ is the wavelength of the seed radiation. $\Delta\nu$ is typically on the order of the Pierce parameter, and outside a narrow frequency bandwidth of order $\rho$ the FEL gain is substantially suppressed. Maintaining accuracy across a detuning range of order $\rho$ is important if one would like to study self-amplified spontaneous emission (SASE), which entails loading the beam with a randomly distributed bunching factor across a small range of frequencies around the resonance. To this end we show in Figure \ref{fig:detuning} the detuning curves obtained from Genesis and from the present approach. In these curves we are in particular plotting the ratio of output power to input power as a function of the detuning of the seed radiation from resonance. Indeed, we reproduce almost exactly the dependence of the total power gain in the undulator on the detuning over the relevant range, although with a slight shift by what appears to be some constant detuning value.

\subsection{\label{subsec:largeonaxis}Off-axis seed}

\begin{figure*}[htb!]
    \centering
    \includegraphics[width=0.65\columnwidth]{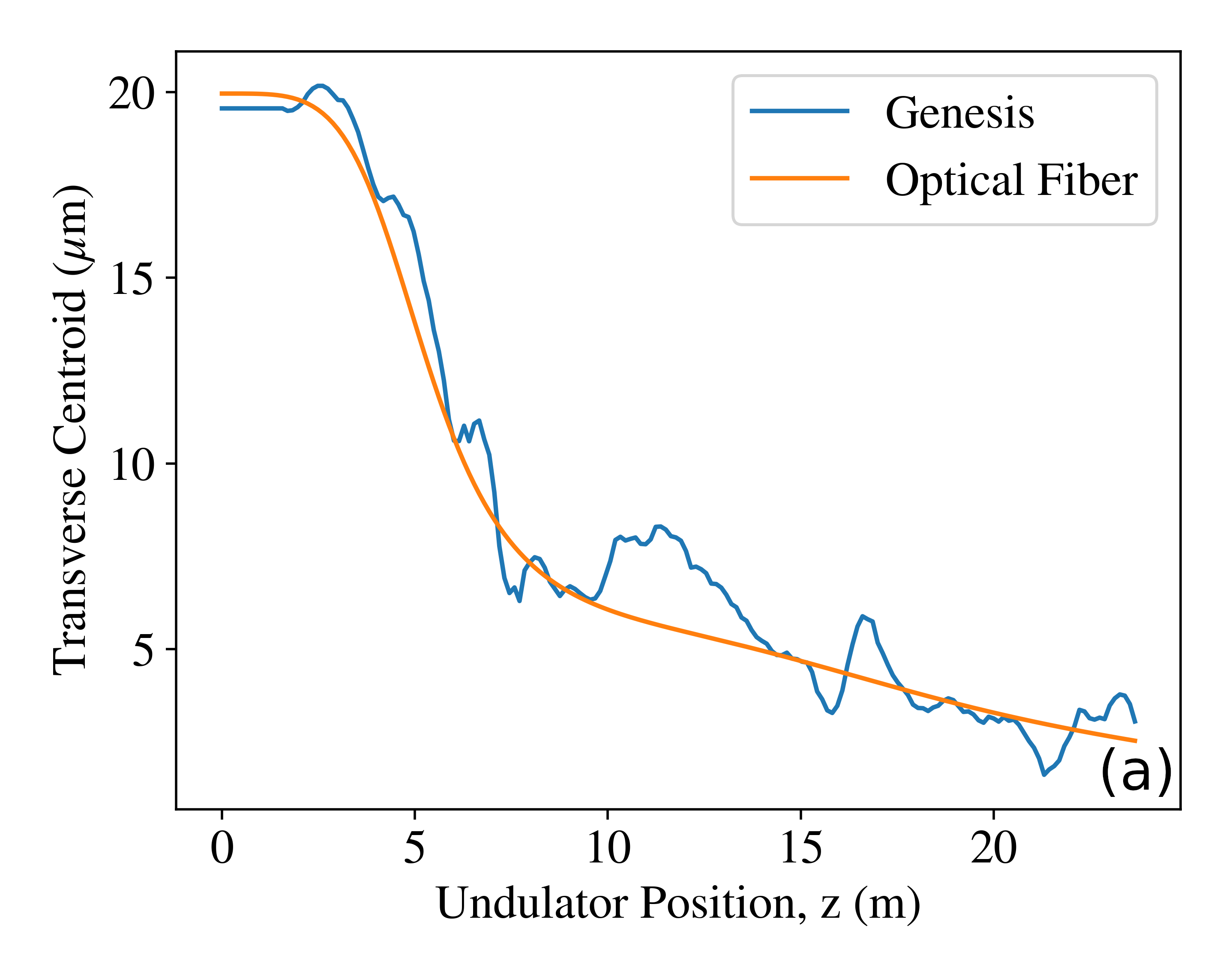}
    \includegraphics[width=0.65\columnwidth]{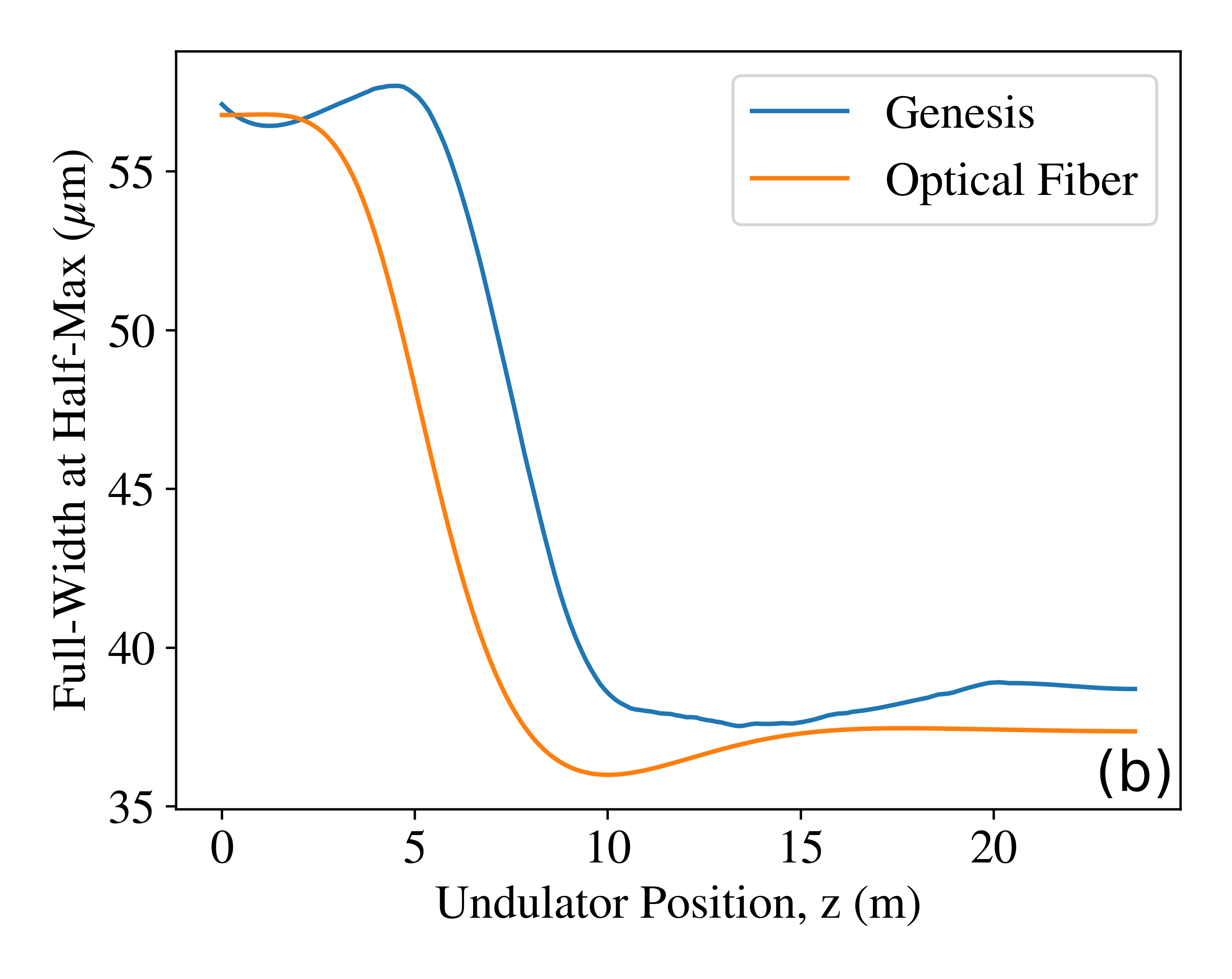}
    \includegraphics[width=0.65\columnwidth]{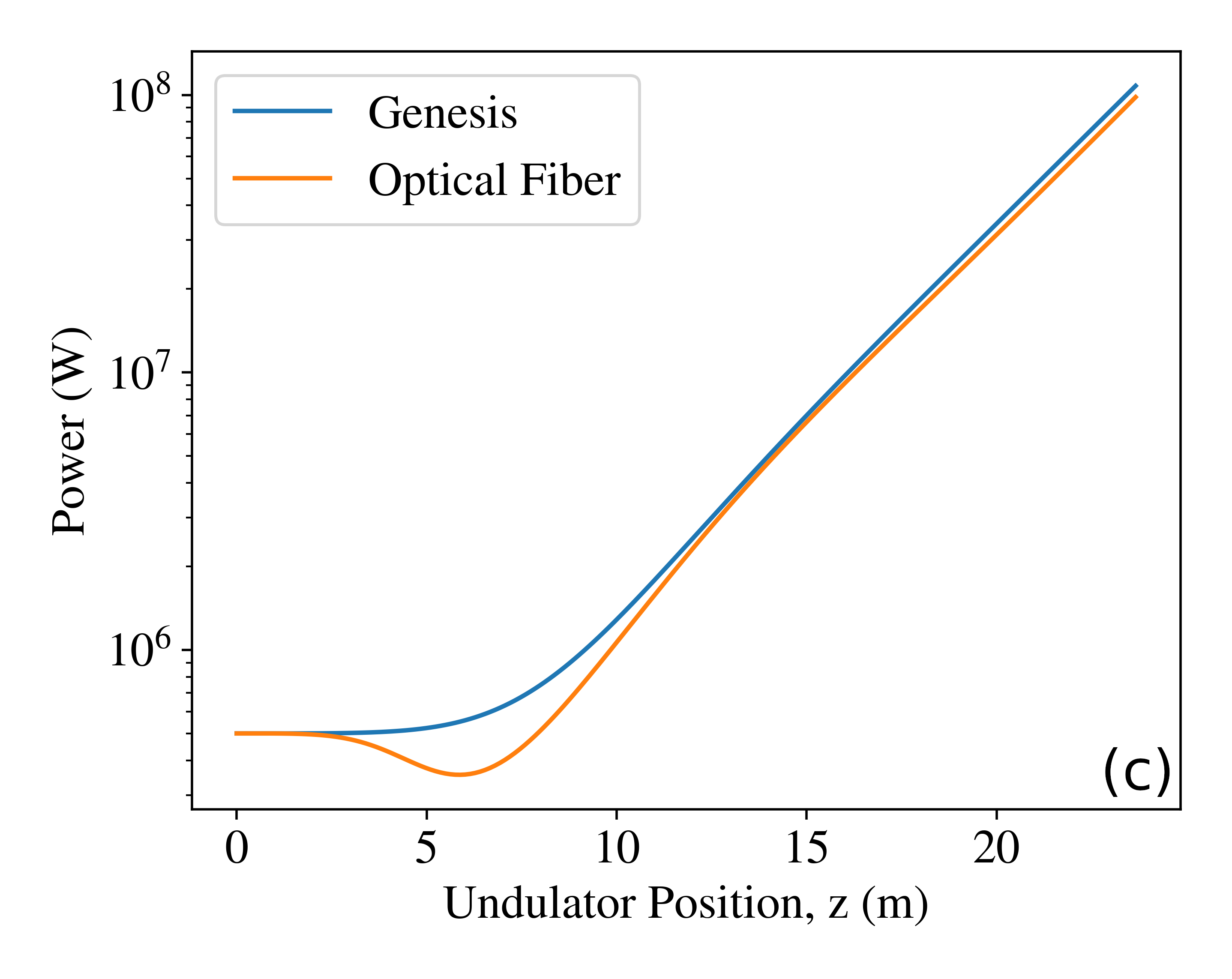}
    \includegraphics[width=0.65\columnwidth]{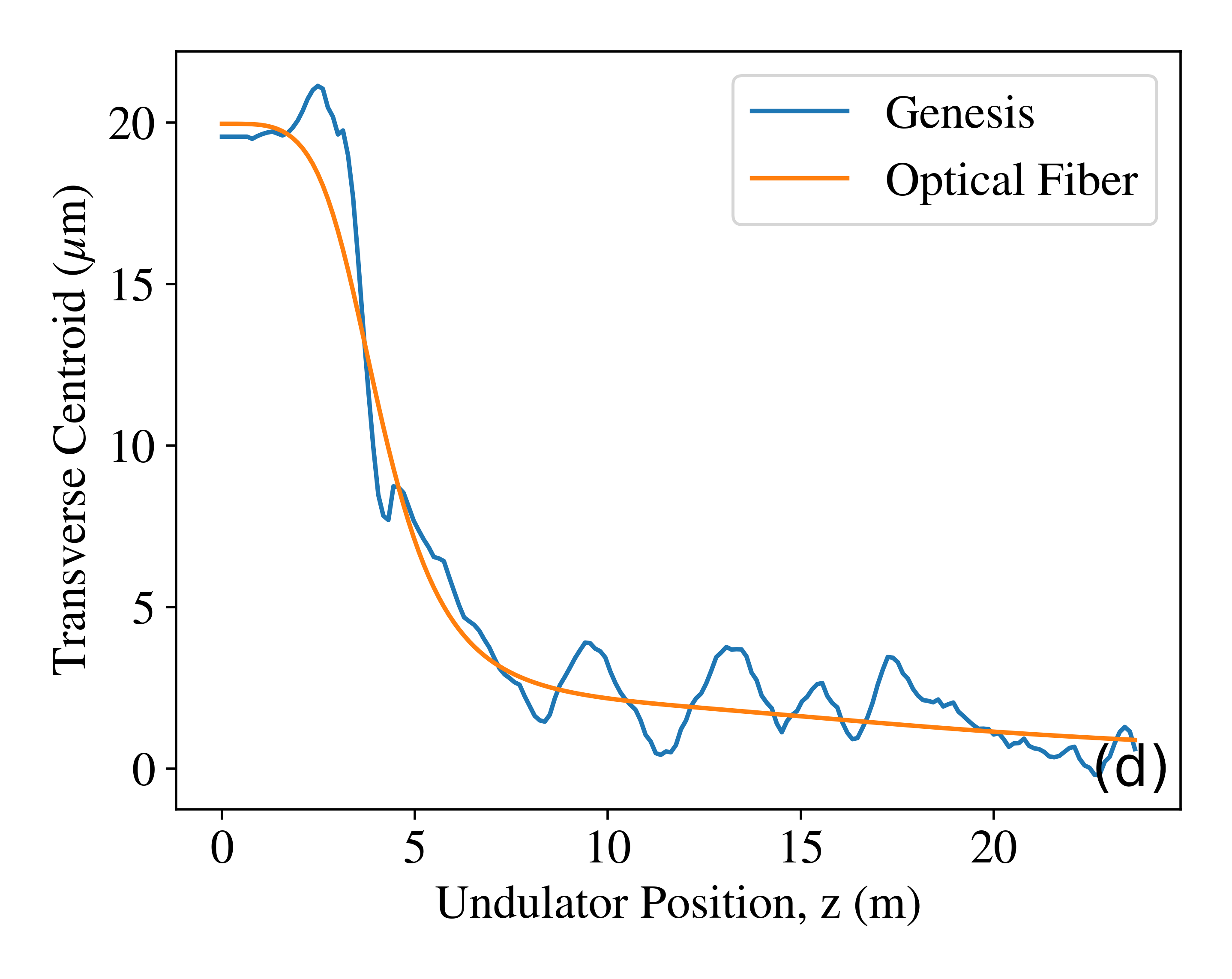}
    \includegraphics[width=0.65\columnwidth]{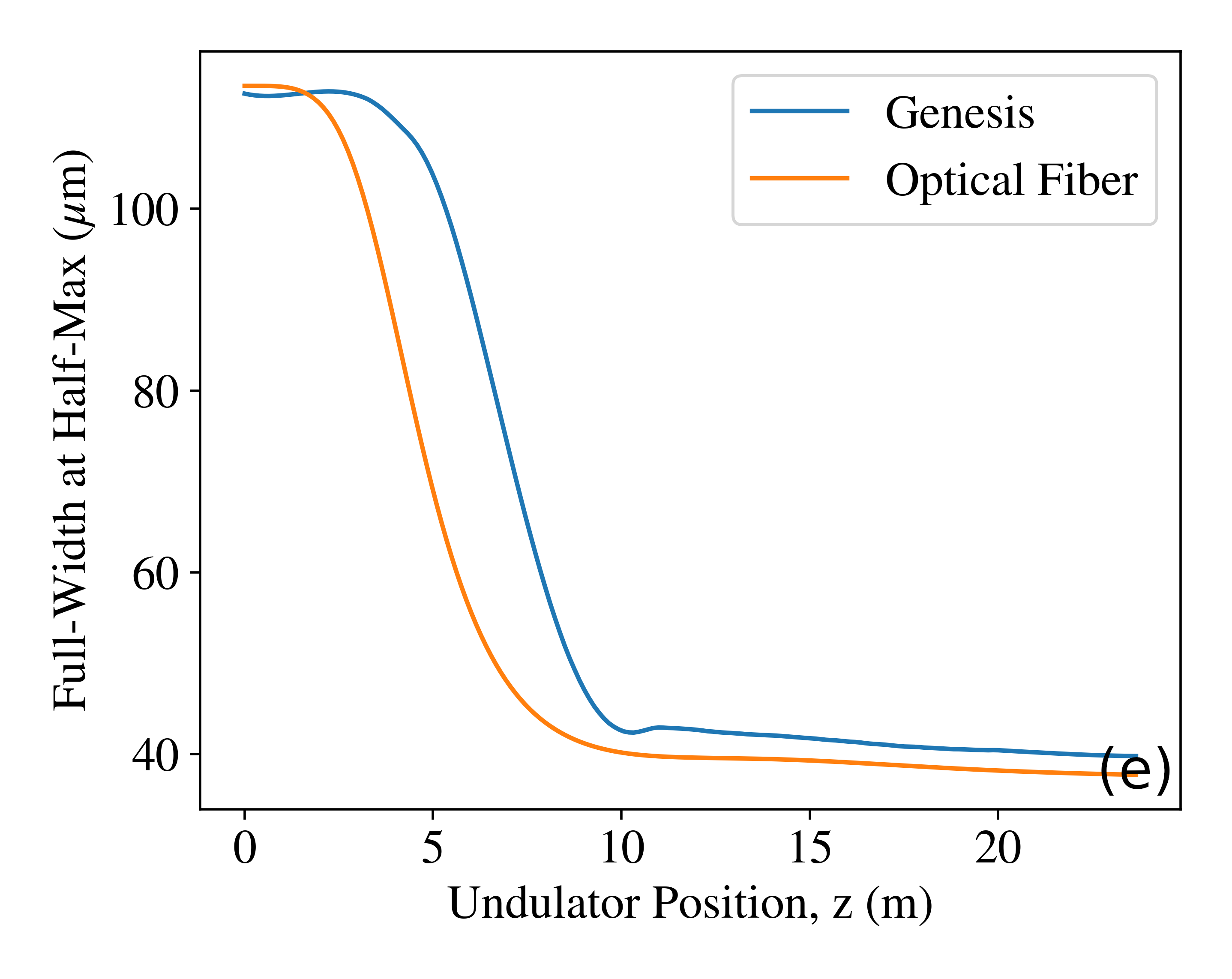}
    \includegraphics[width=0.65\columnwidth]{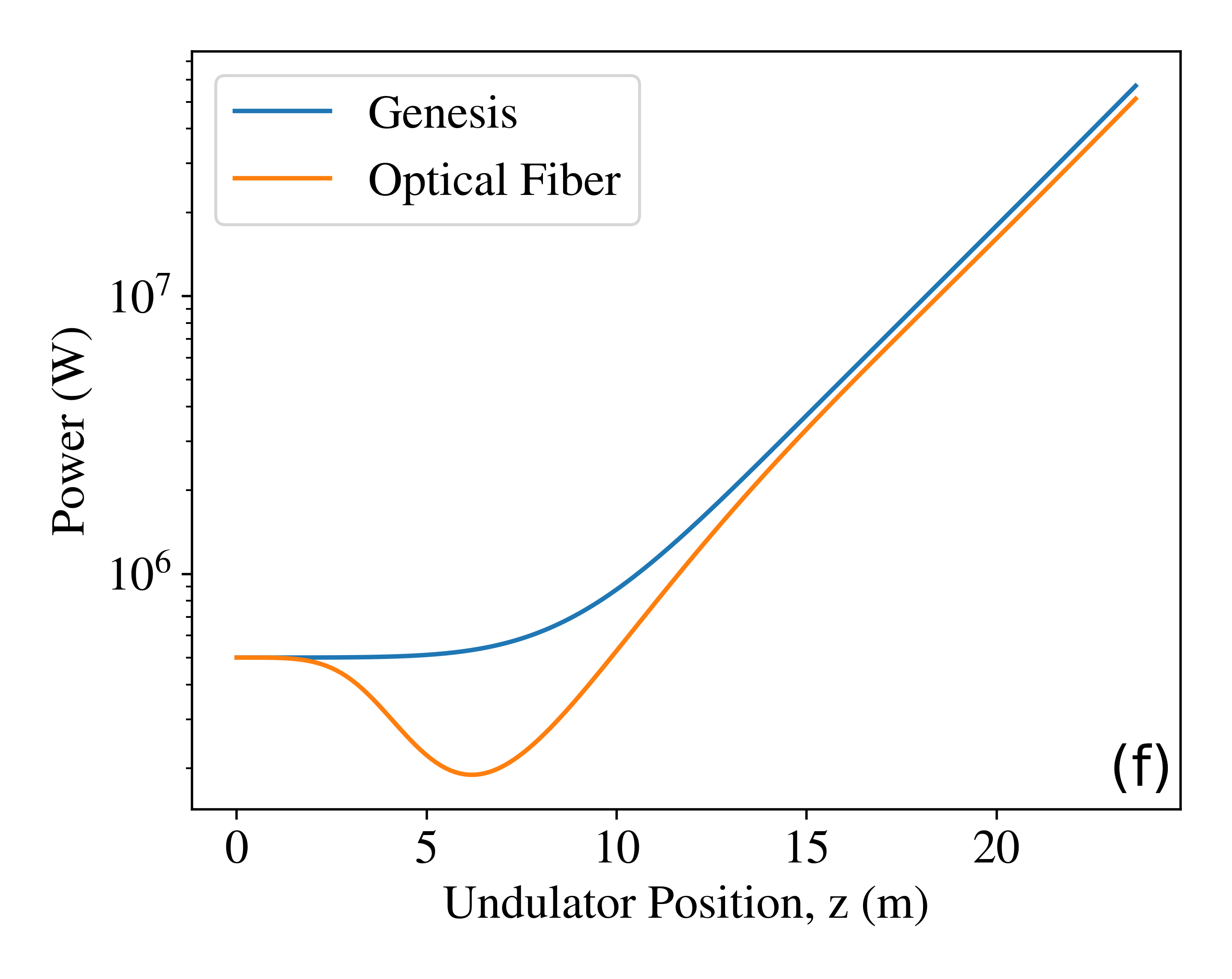}
    \caption{The radiation centroid (a, d), fwhm (b, e) and power growth (c, f) are plotted against the distance along the undulator from Genesis and from the optical fiber theory for two different characteristic input beam sizes. The top (a-c) begin with a 29.8 m Rayleigh length, while the bottom (d-f) begin with a 115.8 m Rayleigh length.}
    \label{fig:smalloffaxis}
\end{figure*}

We will now test the method against an off-axis seed profile, which is of particular interest for seeded FEL amplifiers and x-ray RAFELs. We have used an initial fwhm of 55 $\mu$m and 112 $\mu$m with the initial radiation centroid displaced by $20$ $\mu$m. The results are shown in Figure \ref{fig:smalloffaxis} with an additional plot now showing the transverse centroid evolution. The early disagreement we observed in the last case in the fwhm is again present here, however we again find excellent agreement by the end of the undulator. There is a similar trend in the power, where there is another dip in the early power profile which is largely compensated for by the end of the undulator. The centroid tracking does not seem to suffer from this issue, as the agreement is excellent throughout the undulator. We note the somewhat noisy behavior of the centroid from Genesis is on account of our need to compare against the transverse peak in the intensity profile. The other described metric, the intensity average, is more stable but converges more slowly than the intensity peak because of far-off-axis radiation which is amplified less efficiently on account of the inherent asymmetry in the gain medium when the seed is injected off-axis. This is discussed in more detail in the next section.

\begin{figure}[H]
    \centering
    \includegraphics[width=0.85\columnwidth]{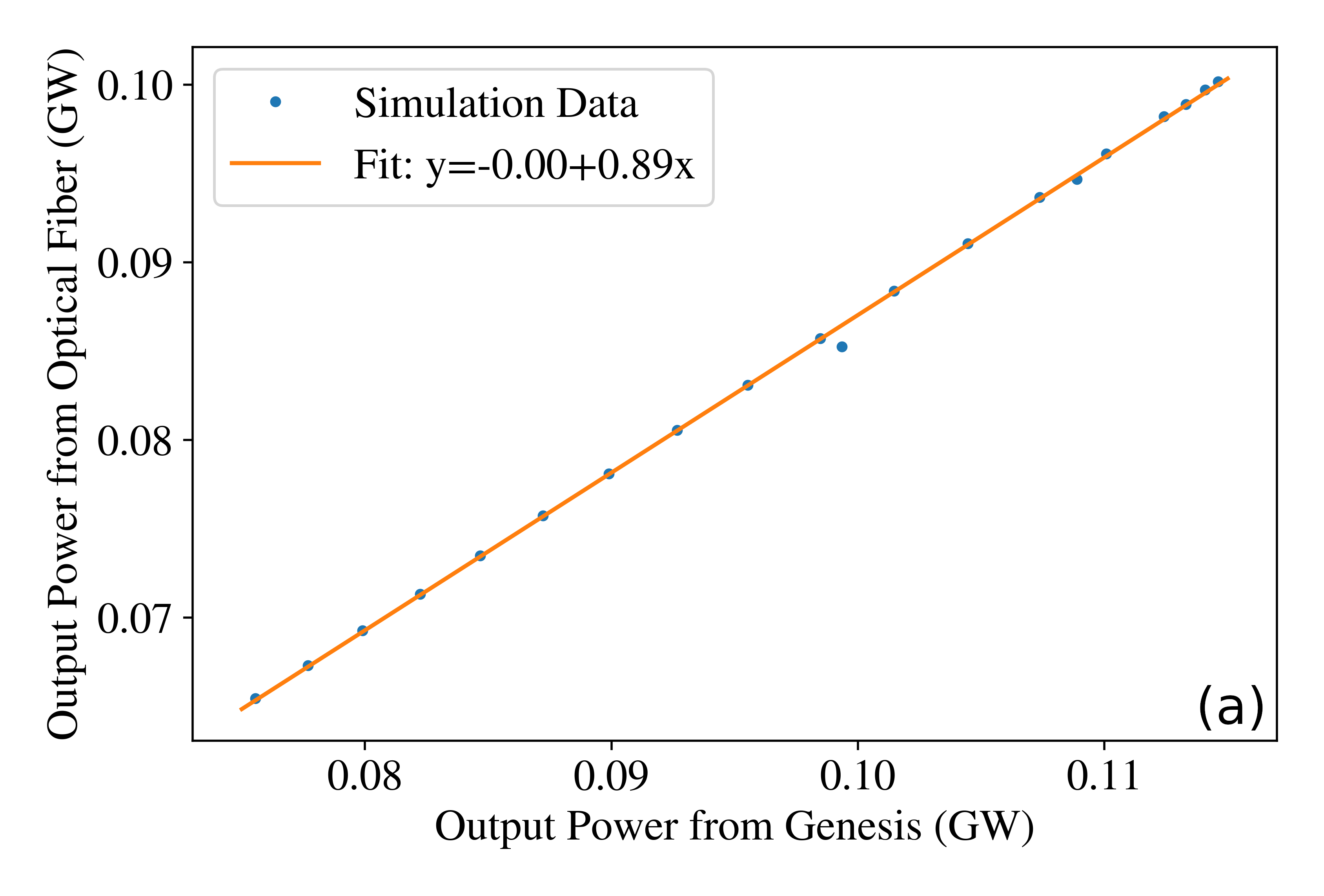}
    \includegraphics[width=0.85\columnwidth]{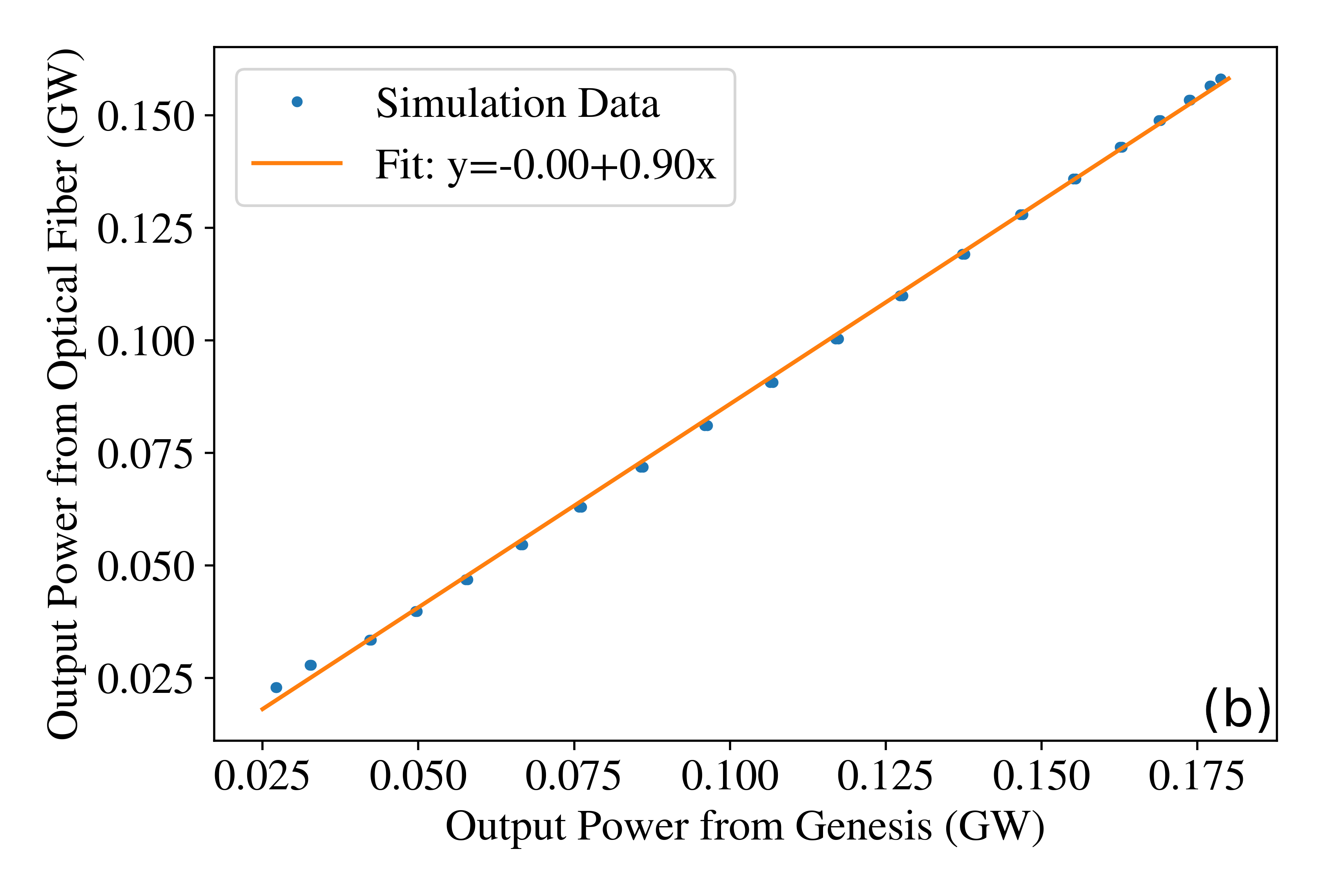}
    \caption{The output power produced from the optical fiber method is plotted against that produced by Genesis as (a) the seed Rayleigh range is scanned from 10 m to 80 m and as (b) the initial radiation offset is varied from -40 $\mu$m to 40 $\mu$m. }
    \label{fig:offaxis_scans}
\end{figure}

Again we demonstrate the broader applicability of the method via a scan over the input radiation parameters in Figure \ref{fig:offaxis_scans}. In both cases we have plotted the output radiation power from the optical fiber approach versus that from Genesis simulations. On the top we have varied the input Rayleigh range from 10 m to 80 m holding the initial offset fixed at 20 $\mu$m and on the bottom we have fixed the Rayleigh range at 28 m and varied the initial offset from $-40$ $\mu$m to $40$ $\mu$m. In both scenarios we obtain promising results similar to those presented in Figure \ref{fig:onaxis_zraylscan}: the dependence is almost perfectly linear with a slope close to unity.

\subsection{\label{subsec:discussion}Discussion of accuracy}

With the apparent accuracy of the method now established, it is worthwhile to take some time to understand precisely in what way the model is faithful to the FEL equations. Any gaussian ansatz for the radiation field will lack some information, because the true FEL field is not a perfect gaussian. Although it is roughly gaussian near the peak of the radiation intensity profile, any radiation energy which is outside of the bounds of the electron beam will free-space diffract away from the axis and experience neither refractive nor gain guiding. This is the reason for our comparison to the fwhm of the radiation mode rather than the rms size: as the edges of the mode diffract away the rms size steadily increases while the fwhm is largely unaffected. This effect is familiar from earlier work \cite{huang2002transverse}. This problem is enhanced in the seeded case, where a large part of the radiation power could in principle lie outside of the bounds of the electron beam. There is a similar effect in the radiation centroid which motivates our decision to compare against the location of peak intensity rather than the average transverse position. It is possible that through similar arguments, we might find better agreement in the field strength by comparing to the peak field intensity rather than the power, however for the purposes of the current work the power suffices. 

\section{\label{sec:jitter}Extension to Electron Beam Trajectory Errors}

In this section we will apply the optical fiber method to understand the effect of a non-ideal electron trajectory on the final radiation profile. Of course, this requires some generalization to the method presented in Section \ref{sec:approach}, in particular to allow for some non-trivial trajectory of the electron beam centroid. We assume that the electron beam centroid is essentially predetermined, primarily by the betatron oscillations associated with the focusing lattice in the FEL, such that we may define the e-beam centroid trajectory $x_{ce}(z)$ with associated angle $p_{x,ce}(z)\equiv x_{ce}'(z)$, as was utilized in \cite{baxevanis2017} from which we borrowed the earlier formalism for the FEL equation \ref{eqn:panoseqn}. Since typically there is jitter in both x and y, which may not be equal, we will keep the method as general as possible by allowing some offset in y as well, with similar definitions of the trajectory.

\subsection{Modifications to the base method}

The mechanics of allowing arbitrary trajectories in x and y are a straightforward extension of the previously presented method. We will discuss here how the equations of Section \ref{sec:approach} must be modified when the electron beam has a non-trivial orbit. To start, the original integral kernel is changed to
\begin{widetext}
\begin{equation}\label{eqn:panoseqnarbtraj}
\begin{split}
    K_1(x,y,p_x,p_y,z,\zeta) =& K_{10}(z,\zeta)\exp\left[-\frac{(p_x-p_{x,ce}(z))^2+(p_y-p_{y,ce}(z))^2}{2k_\beta^2\sigma_x^2}-\frac{(x-x_{ce}(z))^2+(y-y_{ce}(z))^2}{2\sigma_x^2}\right]\\
    &\times\exp\left[-\frac{ik_r(\zeta-z)}{2}\left(p_x^2+p_y^2+k_\beta^2(x^2+y^2)\right)\right],
\end{split}
\end{equation}
\end{widetext}
where we have introduced the predetermined electron beam centroid trajectories $x_{ce}(z)$ and $y_{ce}(z)$, in addition to the centroid momentum trajectories $p_{x,ce}(z)$ and $p_{y,ce}(z)$. In addition to this, we should now perform all Taylor expansions of the refractive index about the centroid of the electron trajectory, since that is now effectively the axis around which the most important physics occurs. We will perform these expansions in both x and y to allow for generally different trajectories in the two planes,
\begin{widetext}
\begin{equation}
    n(x,y,z)^2 = n_0^2(z)+2n_{1x}(z)(x-x_{ce}(z))+2n_{1y}(z)(y-y_{ce}(z))-n_{2x}(z)(x-x_{ce}(z))^2-n_{2y}(z)(y-y_{ce}(z))^2,
\end{equation}
\end{widetext}
This incurs an additional parameter to keep track of, namely $n_{1y}(z)$, but that is not too computationally expensive. Additionally, the definitions of the various components of the refractive index are now modified to read, 
\begin{equation}
\begin{split}
    n_0(z)^2 &= n^2\bigg\rvert_{x=x_{ce}(z),y=y_{ce}(z)},\\
    n_{1x}(z) &= \frac{1}{2}\frac{\partial n^2}{\partial x}\bigg\rvert_{x=x_{ce}(z),y=y_{ce}(z)},\\
    n_{1y}(z) &= \frac{1}{2}\frac{\partial n^2}{\partial y}\bigg\rvert_{x=x_{ce}(z),y=y_{ce}(z)},\\
    n_{2x}(z) &= -\frac{1}{2}\frac{\partial^2 n^2}{\partial^2 x}\bigg\rvert_{x=x_{ce}(z),y=y_{ce}(z)},\\
    n_{2y}(z) &= -\frac{1}{2}\frac{\partial^2 n^2}{\partial^2 y}\bigg\rvert_{x=x_{ce}(z),y=y_{ce}(z)}.
\end{split}
\end{equation}
Finally, with this modified refractive index comes a requisite modified form for the field and the evolution of its mode parameters. First of all, the field should now be represented by
\begin{equation}\label{eqn:gaussianansatz2}
\begin{split}
    E(x,y,z) =& f(z)\exp\left[-\frac{i}{2}Q_x(z)(x-x_0(z))^2\right]\\
    &\times\exp\left[-\frac{i}{2}Q_y(z)(y-y_0(z))^2\right].
\end{split}
\end{equation}
In general, the presence of a non-zero trajectory in y will tend to ``drag" the radiation field off-axis in that direction, hence the new presence of $y_0(z)$ representing the centroid of the radiation field in y. Furthermore, the mode parameters now obey slightly modified differential equations obtained in the same way as before, 
\begin{equation}
\begin{split}
    x_0'(z) =& \frac{k_r}{Q_x(z)}[n_{1x}(z)-n_{2x}(z)(x_0(z)-x_{ce}(z))],\\
    y_0'(z) =& \frac{k_r}{Q_y(z)}[n_{1y}(z)-n_{2y}(z)(y_0(z)-y_{ce}(z))],\\
    Q_x'(z) =& k_rn_{2x}(z)+\frac{Q_x(z)^2}{k_r},\\
    Q_y'(z) =& k_rn_{2y}(z)+\frac{Q_y(z)^2}{k_r},
\end{split}
\end{equation}
and the field amplitude now obeys, 
\begin{widetext}
\begin{equation}
\begin{split}
    f'(z) =& \frac{f(z)}{2k_r}\left[Q_x(z)+Q_y(z)+ik_r^2(-1+n_0^2+(x_0(z)-x_{ce}(z))(2n_{1x}(z)-n_{2x}(z)(x_0(z)-x_{ce}(z)))\right.\\
    &\left.+(y_0(z)-y_{ce}(z))(2n_{1y}(z)-n_{2y}(z)(y_0(z)-y_{ce}(z))))\right].
\end{split}
\end{equation}
\end{widetext}
In addition to all of these expressions, the results of Section \ref{subsec:physicalvalues} are appropriately modified, however these modifications are straightforward.

\subsection{Numerical benchmarks}

\begin{figure*}[htb]
    \centering
    \includegraphics[width=0.65\columnwidth]{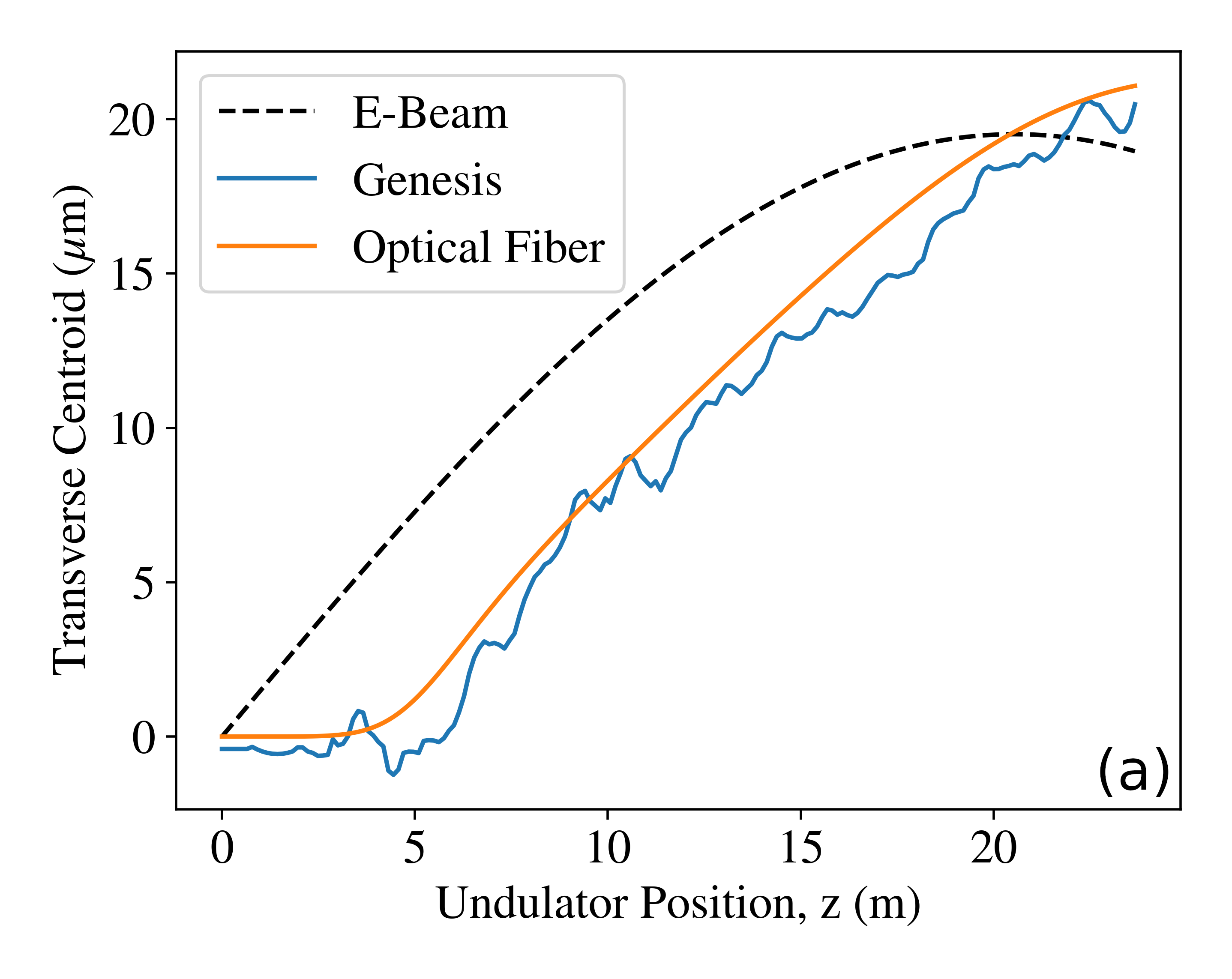}
    \includegraphics[width=0.65\columnwidth]{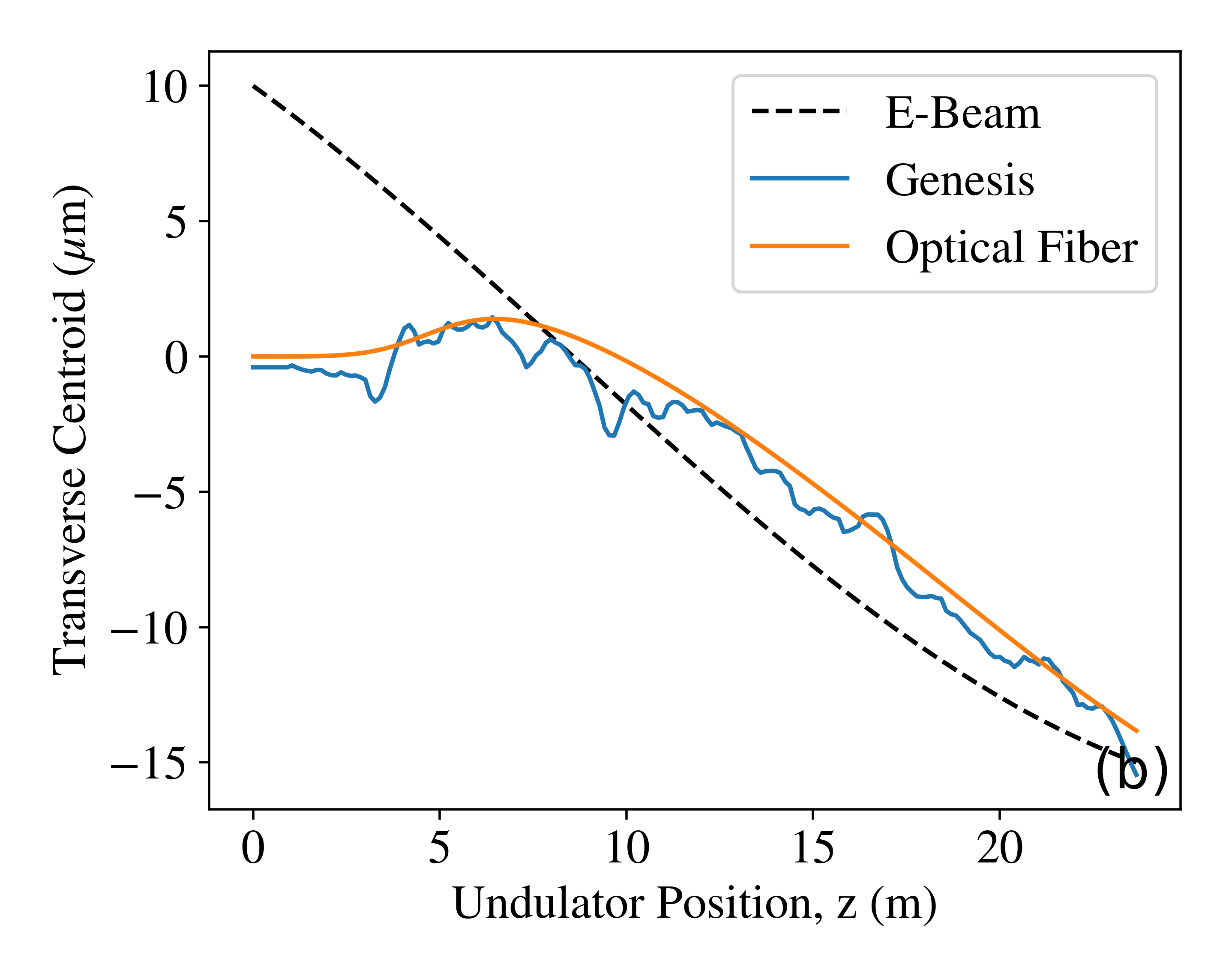}
    \includegraphics[width=0.65\columnwidth]{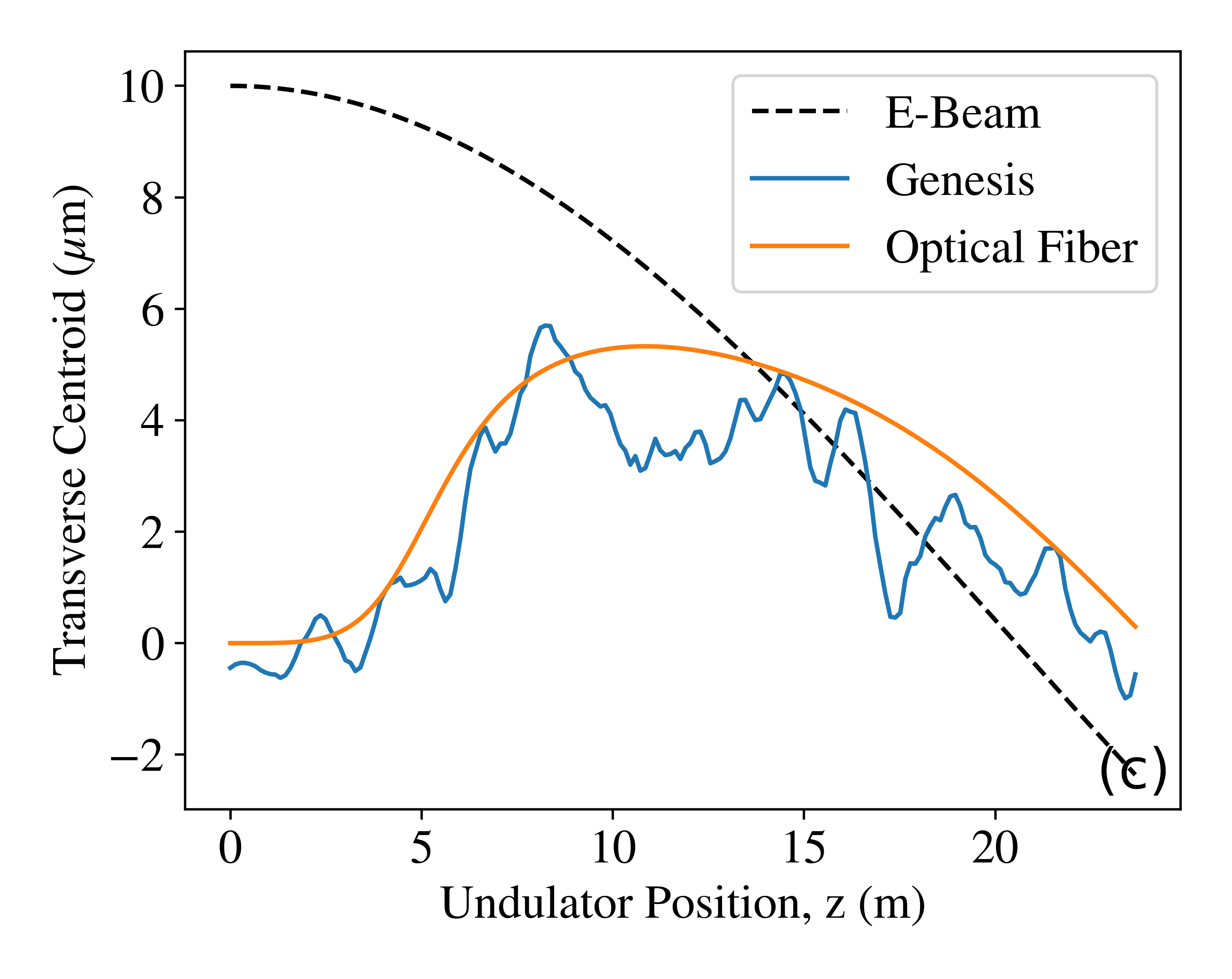}
    \includegraphics[width=0.65\columnwidth]{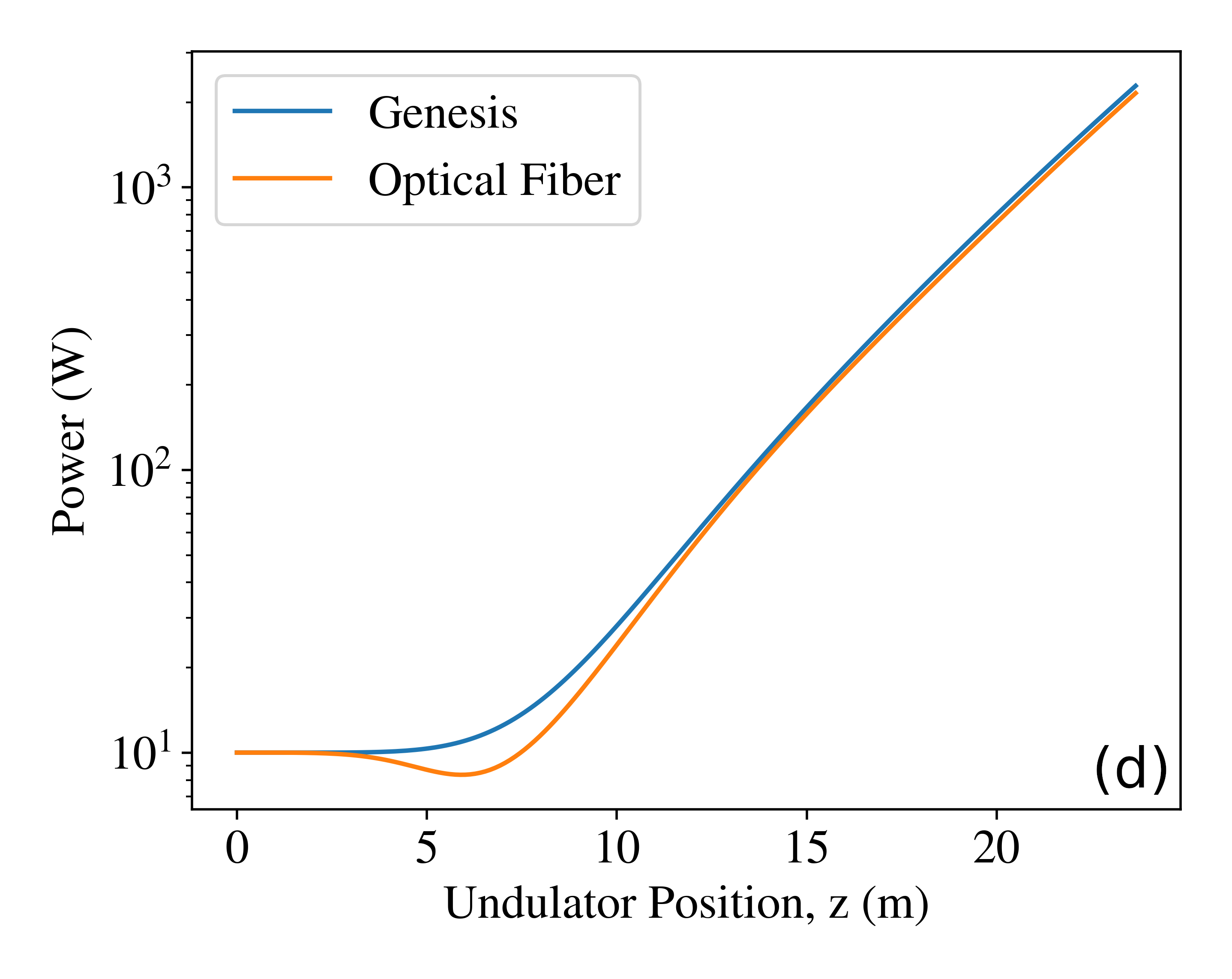}
    \includegraphics[width=0.65\columnwidth]{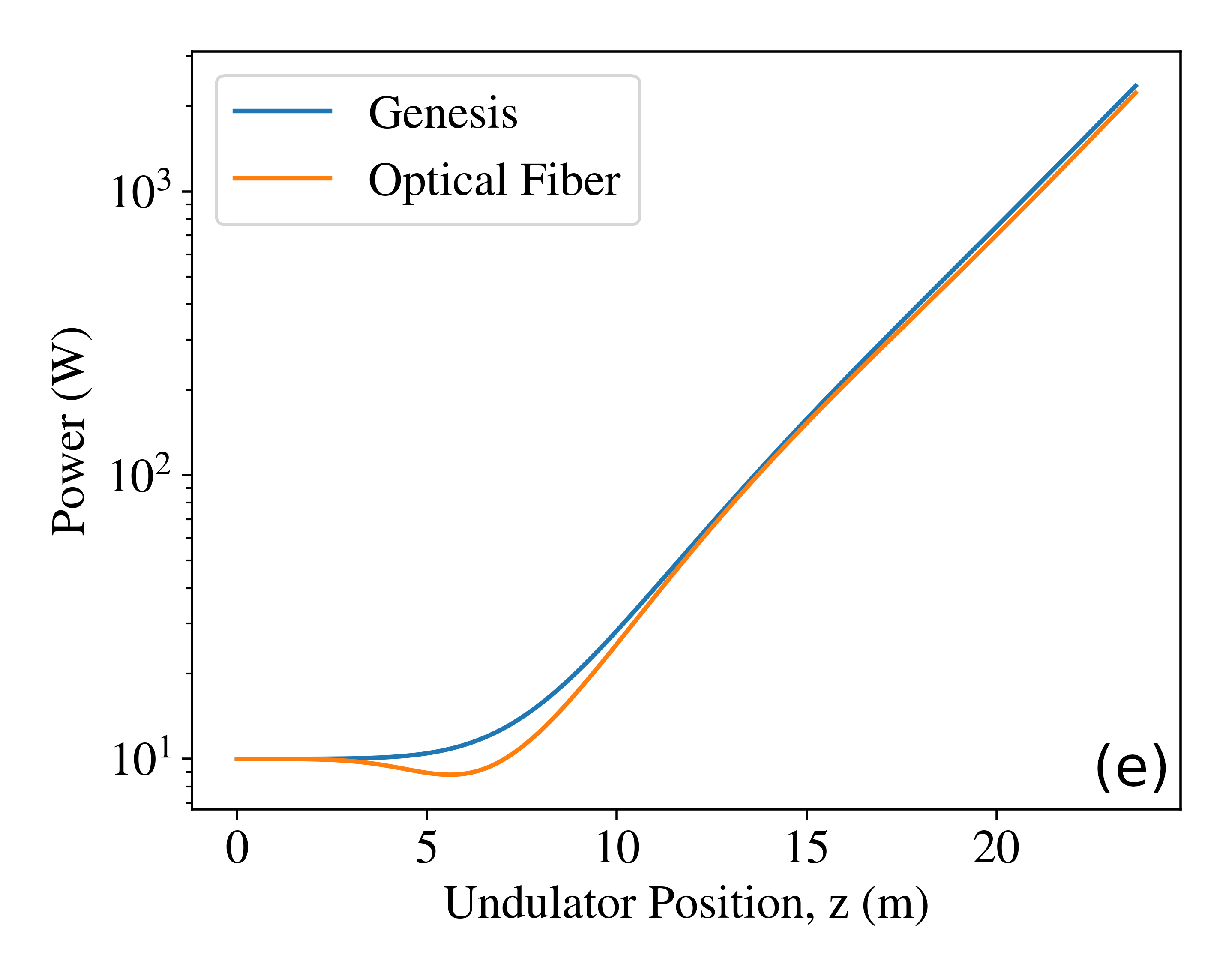}
    \includegraphics[width=0.65\columnwidth]{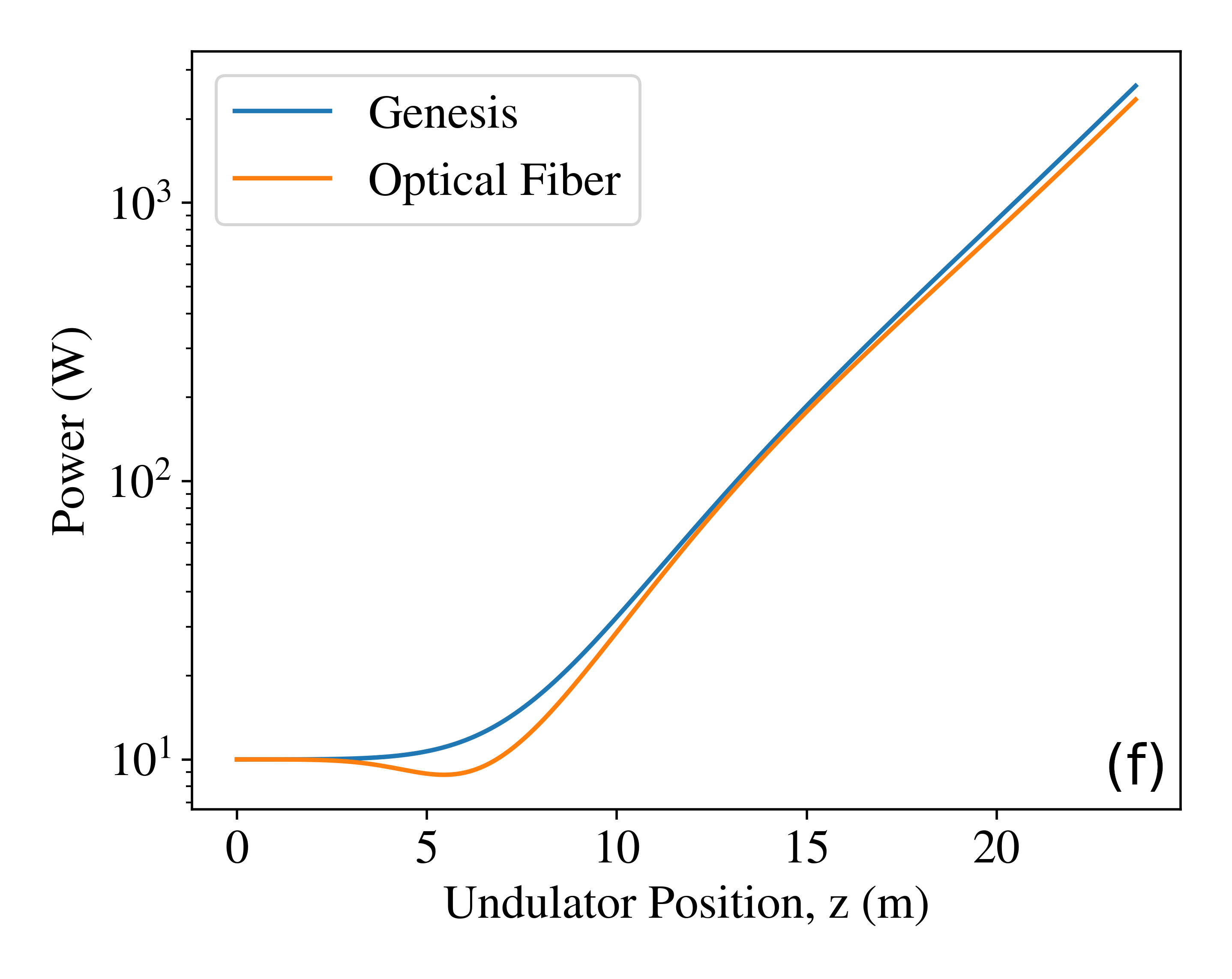}
    \caption{A set of representative plots are shown of the effect of electron beam orbit errors on the FEL pulse. On the top (a-c) the transverse centroid is plotted and on the bottom (d-f) the power as obtained from Genesis and from the optical fiber method. More precisely, these correspond to cases where $x_{ce}(0)=0$ and $p_{x,ce}(0)=1.5$ $\mu$rad (a, d), $x_{ce}(0)=10$ $\mu$m and $p_{x,ce}(0)=-1$ $\mu$rad (b, e), and $x_{ce}(0)=10$ $\mu$m and $p_{x,ce}(0)=0$ (c, f).}
    \label{fig:etraj_benchmarks}
\end{figure*}

Here we demonstrate the accuracy of the method with this additional complication of an arbitrary electron trajectory. The basic FEL parameters used in these benchmarks are the same as those presented in Table \ref{tab:lclsiihe}. As before the simulations employ a 500 kW seed power. Although we have left the electron beam trajectory functions completely arbitrary to this point, in the presence of a smooth focusing lattice, like the one employed in the Genesis simulations, these functions are given by 
\begin{equation}
    x_{ce}(z) = x_{ce}(0)\cos(k_\beta z)+\frac{p_{x,ce}(0)}{k_\beta}\sin(k_\beta z),
\end{equation}
with a similar expression in $y$. The betatron wavenumber $k_\beta$ is equal to the inverse of the average beta function for each plane, $k_\beta=1/\beta_x=\epsilon_x/\sigma_x^2$. We have shown a sample of representative simulations in Figure \ref{fig:etraj_benchmarks}. In the top row we have plotted the transverse centroid of the electron beam as well as that of the radiation beam calculated from Genesis and from the optical fiber method. In the bottom row we show the corresponding comparison in the power development through the undulator. In these representative cases very good agreement is found between the full Genesis simulations and the optical fiber tracking in both the centroids and the power. 

\begin{figure}[htb]
    \centering
    \includegraphics[width=0.85\columnwidth]{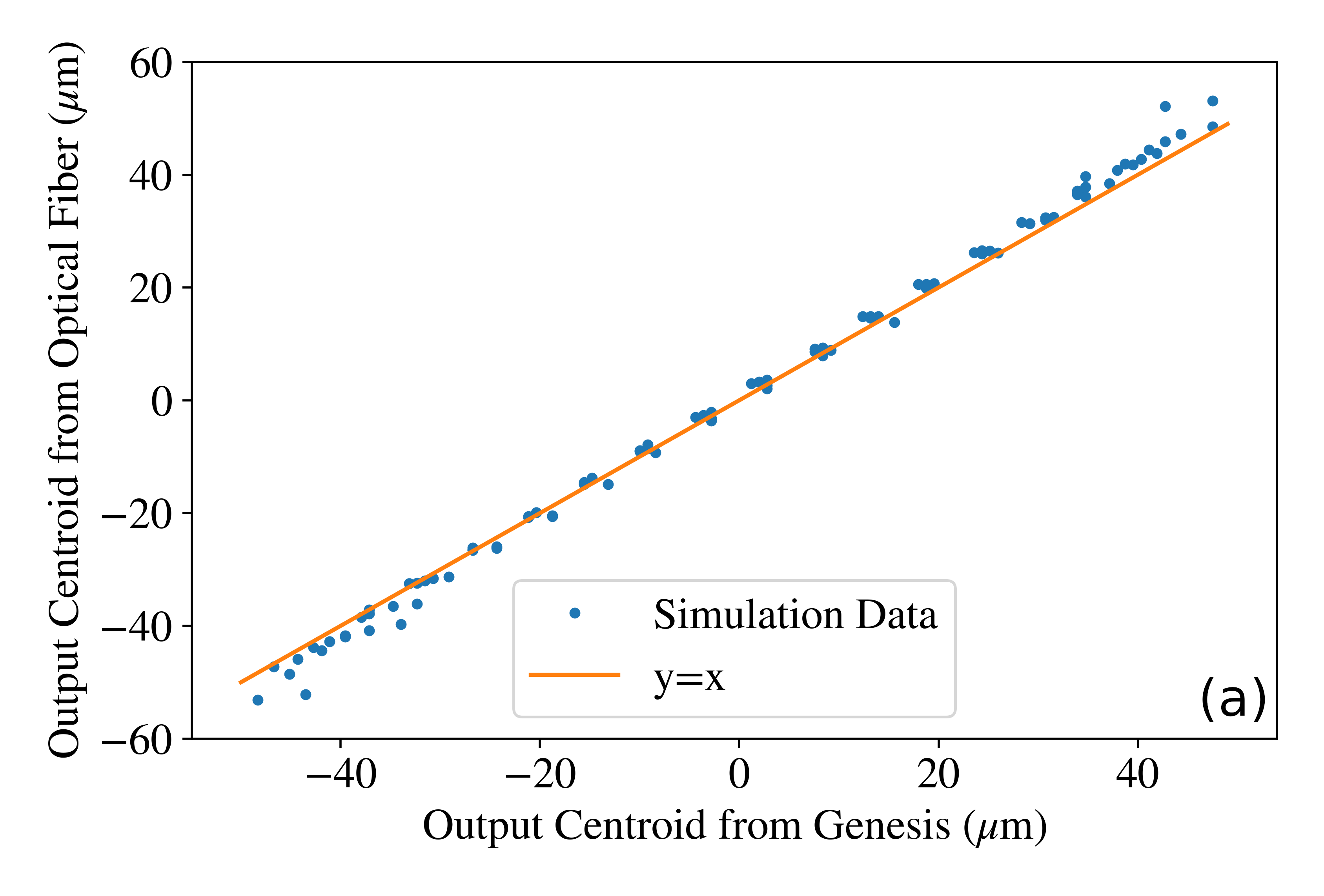}
    \includegraphics[width=0.85\columnwidth]{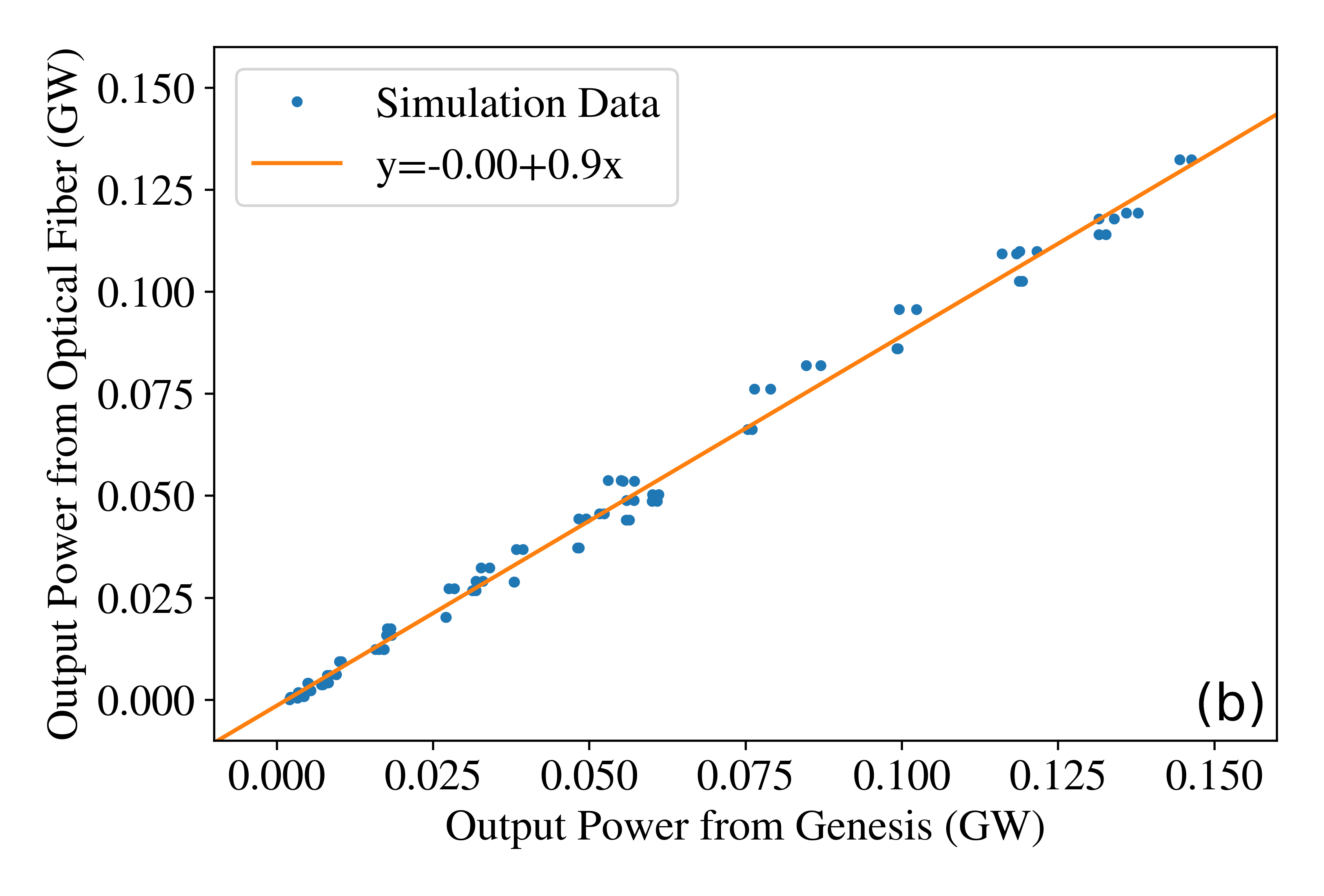}
    \caption{The output radiation centroid (a) and output radiation power (b) obtained from the optical fiber method are plotted against the same values obtained from Genesis with corresponding linear fits. In these plots the initial electron offset and angle are varied from $-20$ $\mu$m to $20$ $\mu$m and from $-4$ $\mu$rad to $4$ $\mu$rad, respectively.}
    \label{fig:etraj_centroids}
\end{figure}

We demonstrate more broadly the utility of the method in Figure \ref{fig:etraj_centroids}. To generate these figures we have performed simulations with amplitudes $x_{ce}(0)$ ranging from $-20$ $\mu$m to $20$ $\mu$m and angular amplitudes $p_{x,ce}(0)$ ranging from -4 $\mu$rad to 4 $\mu$rad. On the top we show the output radiation centroid from the optical fiber approach plotted against that from Genesis with excellent agreement to the expected line $y=x$. On the bottom we show a similar comparison of the output radiation power with the, by now anticipated, linear trend with a slope deviating slightly from unity. 

\subsection{Pointing jitter dependence on electron orbit}

\begin{figure}[htb]
    \centering
    \includegraphics[width=0.85\columnwidth]{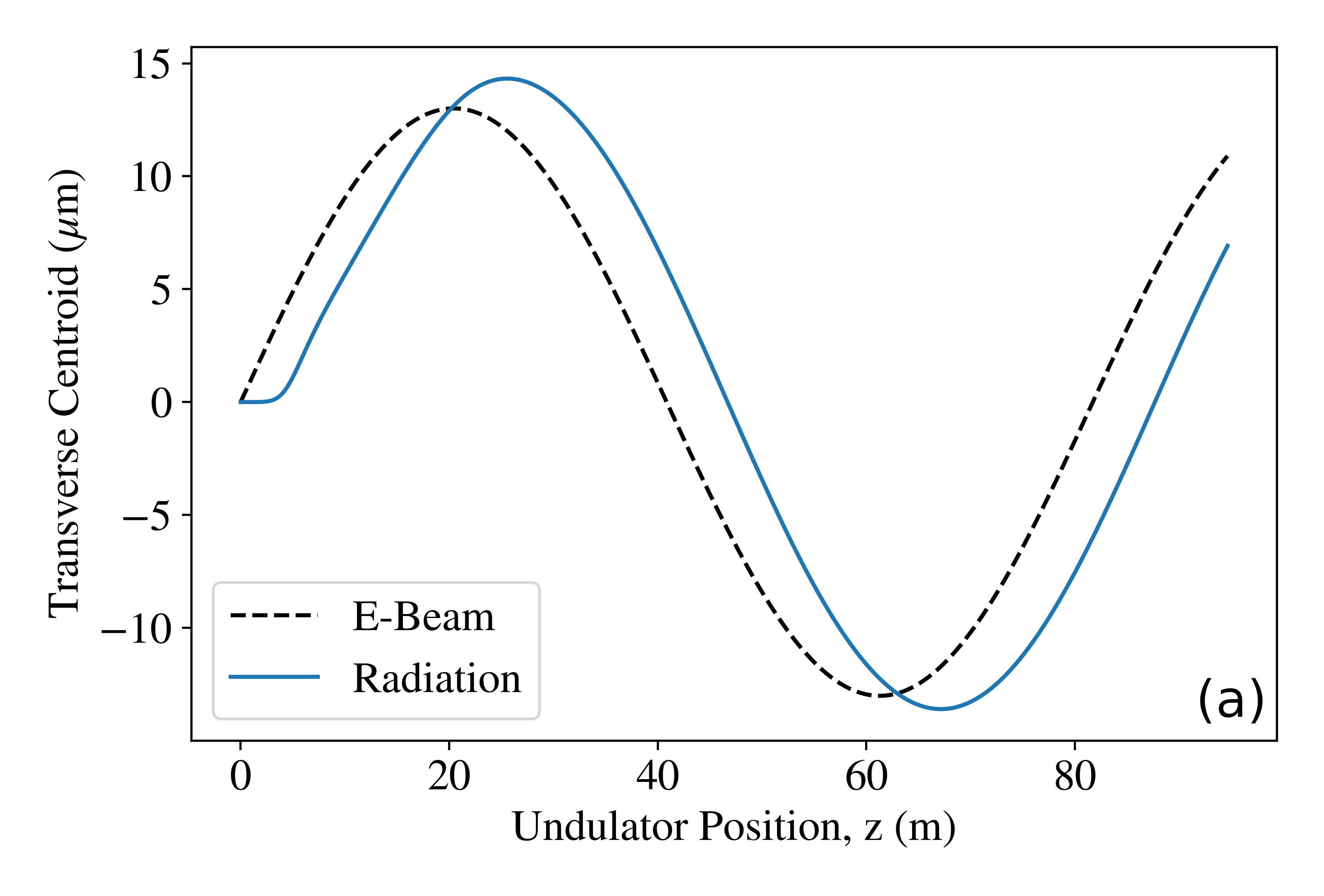}
    \includegraphics[width=0.85\columnwidth]{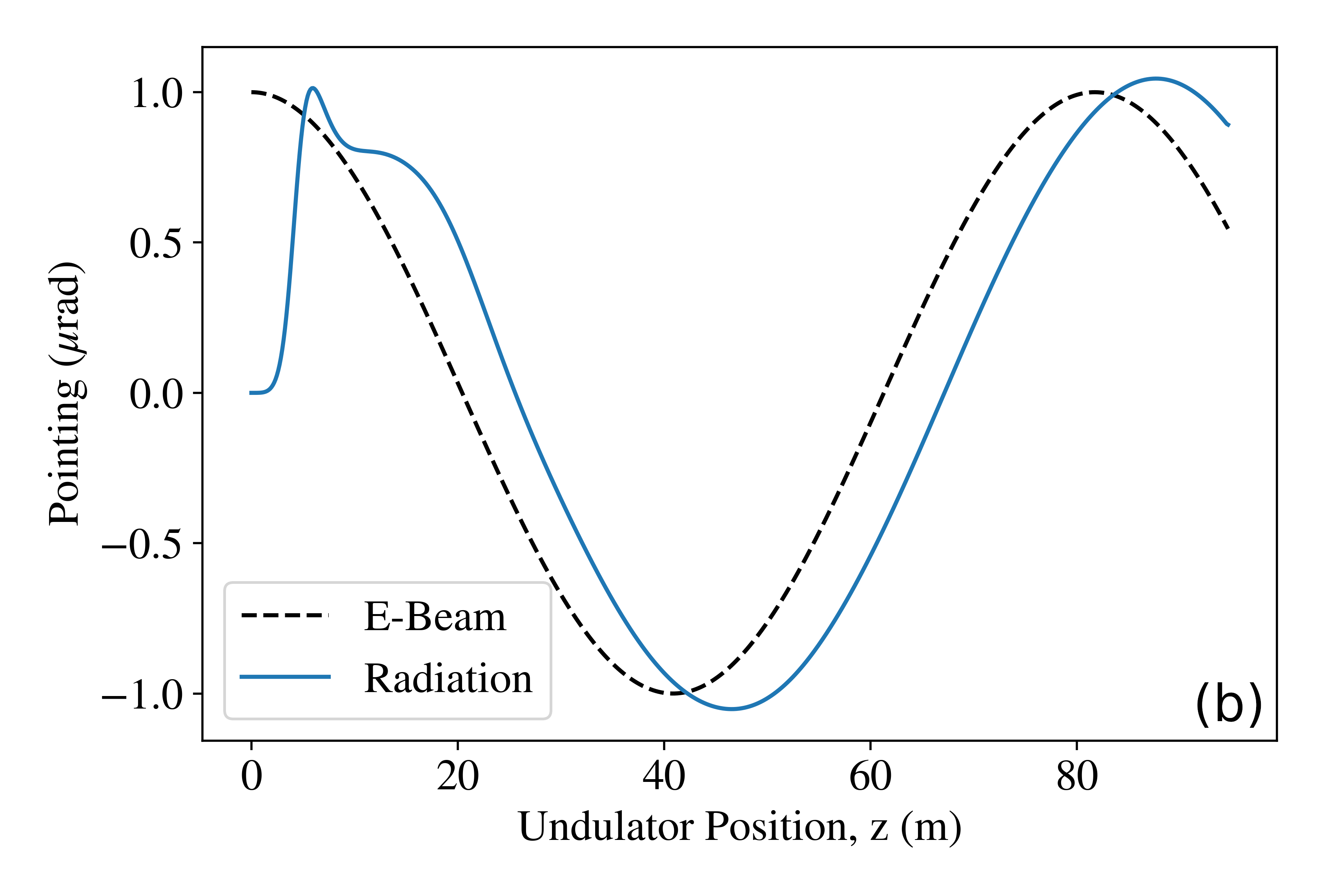}
    \caption{The behavior of the radiation centroid (a) and pointing (b) are shown along with the equivalent parameters for the electron beam along a longer section of undulator which spans roughly one betatron oscillation period.}
    \label{fig:longterm_jitter}
\end{figure}

As was already clear in the numerical benchmarks, the presence of a non-zero electron orbit causes the radiation to be guided along a similar trajectory as the electrons. This behavior is shown from a characteristic case taking place over a distance exceeding a betatron oscillation period in Figure \ref{fig:longterm_jitter} where we have plotted, on the top, the physical transverse centroid and, on the bottom, the angle, or pointing, associated with this centroid oscillation. Here we can see that once the radiation has entered the exponential high-gain regime, its orbit essentially follows that of the electron beam with a position lag which is roughly equal to the length of the lethargy period which precedes the high-gain regime. Furthermore, the radiation beam exhibits a sort of inertia in which it overshoots the peak of the electron beam before turning back around. As the electron orbit becomes more dramatic the perfect sinusoidal nature of the radiation orbit is lost. 

\subsection{Application to SASE FEL operation}

As a matter of practical importance, we would like to show that the method we have described is still able to make useful predictions for SASE FELs which start up from noise in the electron density profile. Although our refractive index model assumes zero initial bunching, we can get around this in one of two ways. The first approach is to re-derive the refractive index without assuming zero initial bunching. This can be done in principle by returning to the methods of \cite{baxevanis2017}, however it is not obvious that the same optical fiber approach could still be employed in the resulting equation. Alternatively, we can simply initialize our seed radiation in such a way that is motivated by the basic understanding of SASE start-up. In particular, we will initialize the radiation profile with the spot size corresponding to the high-gain fundamental mode with an initial power which is uniformly distributed across a broad range of frequency detunings. We expect then that this will come close to the behavior of a SASE FEL after SASE fluctuations are averaged over. 

\begin{figure}[htb]
    \centering
    \includegraphics[width=0.85\columnwidth]{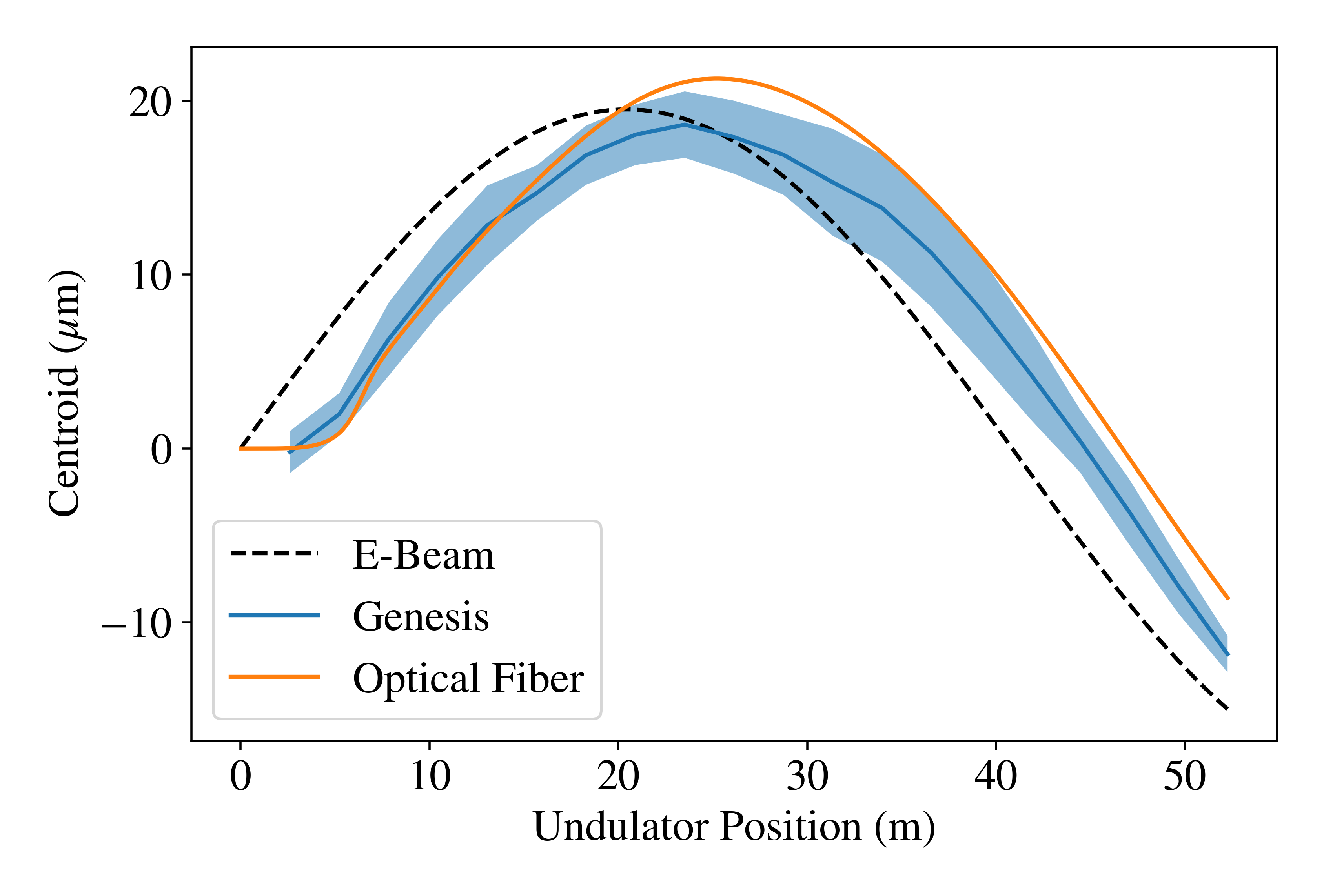}
    \caption{The radiation centroid is plotted as obtained from sixteen independent SASE Genesis simulations and from the optical fiber method alongside the electron beam trajectory.}
    \label{fig:sase_comparison}
\end{figure}

As such, we perform sixteen statistically independent, time-dependent Genesis simulations starting from shot noise and present the average values of the quantities of interest. In particular, we show in Figure \ref{fig:sase_comparison} the radiation centroid behavior in the presence of an initial electron beam angular offset, as determined from SASE runs and from the optical fiber method. The shaded region in this figure indicates a local distance of one standard deviation in the centroid value from the mean at each integration step. The width of this shaded region is determined from two factors: first, the finite number of simulations used in the study, and second the inherent jitter associated with SASE \cite{schneidmiller2017transverse}. The optical fiber method clearly predicts some of the behavior of the radiation centroid in SASE operation, such as the rough length-scale on which the radiation centroid changes, as well as the qualitative shape of the propagation, however it overestimates the maximum offset of the radiation beam from the axis and does not focus as strongly as the SASE beam. This imperfect agreement is unsurprising since we have to utilize an effective seed radiation power in the place of initial bunching, however it is quite close and could be used to make rough estimates. 

\section{\label{sec:conclusions}Conclusions}

We have presented a fast, approximate scheme for solving the fully three-dimensional, single-frequency FEL equations in the linear regime. Our method, which is based on treating the FEL gain medium like a parabolic optical fiber, provides a highly accurate depiction of the radiation mode development in an FEL even in the presence of various non-ideal effects such as transverse offsets in the x-ray beam and orbit errors in the electron beam. The agreement of our approach in tracking the radiation centroid in the presence of these non-ideal effects is of particular note. Although the method we have explicitly considered here is based on a second-order Taylor expansion around the electron beam orbit, we have presented the concept in sufficiently general detail so that different second-order approximations to the FEL refractive index can be considered under the same general framework. 

Two applications stand out as ideal for this particular method: the regenerative amplifier FEL and FEL pointing jitter studies. The application of this approach to RAFEL tolerance and optimization studies will be the subject of a forthcoming publication. Additional applications may be envisioned, in particular any applications for which the FEL process is seeded. Notable examples include the double-bunch FEL \cite{emma2017compact,halavanau2019very}, self-seeding \cite{amann2012demonstration}, or the fresh-slice FEL architecture \cite{lutman2016fresh}. One can also imagine an extension of the method to the tracking of non-gaussian mode profiles. In principle the same general principles can be applied to any field profile which solves Eq. \ref{eqn:dielectricwaveguideeqn} with a particular form of the refractive index as long as the FEL index is appropriately approximated to that particular form. The most obvious extension of this form is to Laguerre-Gaussian modes, which also solve the dielectric waveguide wave equation with a parabolic refractive index.

Although the method thus far applies only in the linear regime prior to saturation, one can imagine ways in which the effects of saturation can be captured at least at its onset. In particular, appropriate application of conservation laws \cite{bonifacio1980coherent, hemsing2020simple} or quasilinear FEL theory \cite{vinokurov2001quasilinear} are standard approaches to including the early effects of saturation. In addition to this, an extension to tapered FELs, both in the linear and saturated regimes, would be highly relevant for the future direction of the RAFEL, and more generally for high power seeded FELs. 

\section*{Acknowledgements}

This work was supported by the Department of Energy, Laboratory Directed Research and Development program at SLAC National Accelerator Laboratory, under contract DE-AC02-76SF00515. R.R.R. acknowledges support from the William R. Hewlett graduate fellowship through the Stanford Graduate Fellowship (SGF) program. 

\appendix

\section{\label{app:index}Analytic expression for the FEL refractive index with a gaussian mode}

In this section we will give the analytic form of the FEL refractive index for a gaussian mode and arbitrary electron trajectory. Since the impact of electron trajectory jitter is in some sense an additional impact on top of x-ray trajectory offsets, in particular in the case of the RAFEL, we will explicitly separate out their impacts. The end result is of the form
\begin{widetext}
\begin{equation}
    n^2(x,y,z,\zeta) = 1+\frac{4\pi k_\beta^2\sigma_x^2}{k_r}\int_0^zd\zeta\frac{f(\zeta)}{f(z)}K_{10}(z,\zeta)\sqrt{\frac{1}{g_x(z,\zeta)g_y(z,\zeta)}}e^{h_{0}(x,z,\zeta)+h_{ce}(x,z,\zeta)+h_{0}(y,z,\zeta)+h_{ce}(y,z,\zeta)},
\end{equation}
\end{widetext}
where $g_x(z,\zeta)=i+k_rk_\beta^2\sigma_x^2(z-\zeta)-\sigma_x^2Q_x(\zeta)\sin(k_\beta(z-\zeta))^2$ with a similar expression for y. The arguments of the exponential take the form
\begin{widetext}
\begin{equation}
\begin{split}
    h_{0}(x,z,\zeta) =& \frac{i}{2}\left[ Q_x(z)(x-x_0(z))^2 + \frac{A(z,\zeta)\left[x^2A(z,\zeta)-\sigma_x^2Q_x(\zeta)(x^2-2xx_0(\zeta)\cos(k_\beta(z-\zeta))+x_0(\zeta)^2)\right]}{\sigma_x^2\left(A(z,\zeta)-\sigma_x^2Q_x(\zeta)\sin(k_\beta(z-\zeta))^2\right)} \right],\\
    h_{ce}(x,z,\zeta) =& \frac{p_{x,ce}(z)\left[ip_{x,ce}(z)+2k_\beta\sigma_x^2Q_x(\zeta)\sin(k_\beta(z-\zeta))\left(x_0(\zeta)-x\cos(k_\beta(z-\zeta))\right)\right]}{2k_\beta^2\sigma_x^2\left(A(z,\zeta)-\sigma_x^2Q_x(\zeta)\sin(k_\beta(z-\zeta))^2\right)}\\
    &-\frac{p_{x,ce}(z)^2}{2k_\beta^2\sigma_x^2}-\frac{x_{ce}(z)^2-2xx_{ce}(z)}{2\sigma_x^2} ,
\end{split}
\end{equation}
\end{widetext}
where in each of these $A(z,\zeta)=i+k_rk_\beta^2\sigma_x^2(z-\zeta)$. In these forms, $h_{ce}$ vanishes in the absence of a non-zero electron beam trajectory. When $x_0(z)$ also vanishes, the ideal FEL scenario without x-ray offset is retrieved. 

\section{\label{app:panoscomparison}Comparison to other methods}

To the authors' knowledge, there is one other fast approximate method for solving the 3D FEL equations in the time-independent limit, described in \cite{baxevanis2017} which we will refer to as the Mode Expansion Method. This approach also takes a gaussian ansatz for the radiation field and can generally allow for a radiation profile which is off-axis or at an angle, which implies that it can also be thought of as a second-order expansion of an effective refractive index. The difference between our method and this lies in the way in which the refractive index expansion is taken. While we have chosen to prioritize the near-axis physics by taking a simple Taylor-expansion around the origin, the authors of \cite{baxevanis2017} enforce the numerical solution's accuracy over a larger transverse range by implicitly using more complicated expressions which are weighted by the local value of the field. These methods compare well to the simulation rms output as a result, however do not compare as well to the fwhm output. In other words, they give up accuracy near the undulator axis and in return are more accurate over the full transverse range in an averaged or root-mean-square sense. 

\begin{figure}[htb]
    \centering
    \includegraphics[width=0.85\columnwidth]{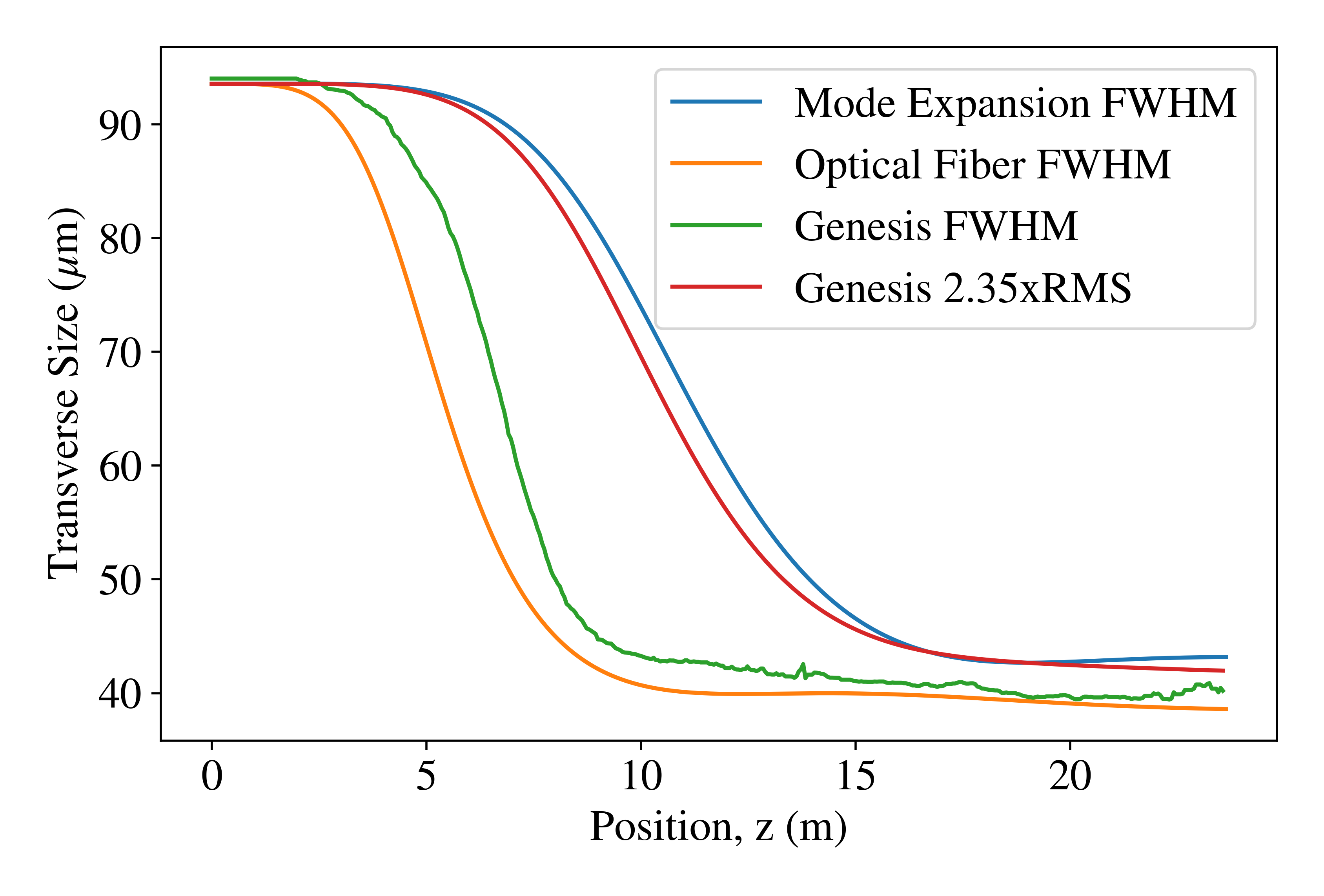}
    \caption{The radiation fwhm is shown as calculated from: the mode expansion approach, the optical fiber approach, Genesis field file outputs, and the equivalent fwhm for a gaussian beam with the rms size provided by Genesis. }
    \label{fig:modeexpansioncomparison}
\end{figure}

We include Figure \ref{fig:modeexpansioncomparison} to clarify this point. As we have discussed earlier, the radiation field of the FEL is not truly gaussian as a result of radiation diffracting away from the axis if it is not well-overlapped with the electron beam. As a result, one expects that the fwhm and rms sizes of the radiation profile will not be related by the usual formula for a gaussian function fwhm $=2\sqrt{2\log(2)}$rms $\approx 2.35$rms. Further, we will expect that while our method will explicitly reproduce the fwhm calculated from the Genesis result, the Mode Expansion Method will yield a gaussian approximation to the simulation result which shares the same rms size. Therefore, we expect that a direct comparison of the Genesis fwhm to our calculated fwhm will be in close agreement, while a direct comparison of the Genesis effective gaussian fwhm (the Genesis rms size times 2.35) will agree well with the fwhm predicted by the Mode Expansion method. 

Indeed, this is exactly what we observe in Figure \ref{fig:modeexpansioncomparison}. In this figure we show the fwhm of the radiation as calculated four ways: from the referenced mode expansion numerical method, from our optical fiber method, explicitly calculated from the field output of Genesis, and finally calculated using the gaussian beam formula fwhm $=2\sqrt{2\log(2)}$rms $\approx 2.35$rms with the rms size taken from Genesis. As we discussed above, the explicit calculation of the fwhm using the Genesis field files agrees very well with our optical fiber approach, while the equivalent fwhm of a gaussian beam with rms size equal to that in the simulations agrees best with the mode expansion method.

\bibliography{apssamp}

\end{document}